\documentclass[preprintnumbers, superscriptaddress, showpacs, nofootinbib, amsfonts, amsmath, amssymb, aps, prd, notitlepage, floatfix]{revtex4-2}

\usepackage[utf8]{inputenc}
\usepackage{amsmath,amssymb,amsfonts,mathrsfs,bbold}
\usepackage{yfonts}
\usepackage{graphicx}
\usepackage[colorlinks=true,citecolor=blue,linkcolor=blue,urlcolor=blue]{hyperref}
\usepackage[dvipsnames]{xcolor}
\usepackage{lipsum}
\usepackage{slashed}
\usepackage[normalem]{ulem}
\usepackage{url}
\usepackage{physics}

\usepackage{amssymb, bm, amsmath, graphicx}
\usepackage{enumerate}
\usepackage{amsfonts}
\usepackage{mathtools}
\usepackage{latexsym}
\usepackage{epstopdf}
\usepackage{subfigure}
\usepackage{tikz}
\usetikzlibrary{decorations.pathmorphing, decorations.markings, shapes.geometric, shapes.misc, calc, math}
\tikzstyle{gluon}=[decorate, decoration={coil,aspect=0.8, amplitude=1.5pt,  segment length=3pt}]
\usepackage{stackrel}
\usepackage[most]{tcolorbox}

\def\eq#1{{Eq.~(\ref{#1})}}
\def\fig#1{{Fig.~\ref{#1}}}
\newcommand{\ben}{\begin{eqnarray*}}
\newcommand{\een}{\end{eqnarray*}}

\newcommand{\rvec}[1]{\overrightarrow{#1}}
\newcommand{\lvec}[1]{\overleftarrow{#1}}

\newcommand{\as}{\alpha_s}

\newcommand{\ord}[1]{\mathcal{O} \left( #1 \right)}

\allowdisplaybreaks

\begin{document}

\title{First study of polarized proton-proton scattering with small-$x$ helicity evolution}

    \makeatletter  
\def\@fnsymbol#1{\ensuremath{\ifcase#1\or *\or \dagger\or \ddagger\or
   \mathsection\or \mathparagraph\or \|\or **\or \dagger\dagger
   \or \ddagger\ddagger \or \mathsection\mathsection \else\@ctrerr\fi}}
    \makeatother
\author{Daniel~Adamiak }
\affiliation{Jefferson Lab, Newport News, Virginia 23606, USA}
\author{Nicholas~Baldonado}
\affiliation{Department of Physics, New Mexico State University, Las Cruces, New Mexico 88003, USA}
\author{Yuri~V.~Kovchegov}
\affiliation{Department of Physics, The Ohio State University, Columbus, Ohio 43210, USA}
\author{Ming~Li}
\affiliation{Department of Physics, The Ohio State University, Columbus, Ohio 43210, USA}
\author{W.~Melnitchouk}
\affiliation{Jefferson Lab, Newport News, Virginia 23606, USA}
\author{Daniel~Pitonyak }
\affiliation{Department of Physics, Lebanon Valley College, Annville, Pennsylvania 17003, USA}
\author{Nobuo~Sato}
\affiliation{Jefferson Lab, Newport News, Virginia 23606, USA}
\author{Matthew~D.~Sievert}
\affiliation{Department of Physics, New Mexico State University, Las Cruces, New Mexico 88003, USA}

\author{Andrey~Tarasov}
\affiliation{Department of Physics, North Carolina State University, Raleigh, North Carolina 27695, USA}
\affiliation{Joint BNL-SBU Center for Frontiers in Nuclear Science (CFNS) at Stony Brook University, Stony Brook, New York 11794, USA}
\author{Yossathorn~Tawabutr}
         \affiliation{\mbox{Department of Physics, University of Jyv\"askyl\"a,  P.O. Box 35, 40014 University of Jyv\"askyl\"a, Finland}}
         \affiliation{Helsinki Institute of Physics, P.O. Box 64, 00014 University of Helsinki, Finland \\
      \vspace*{0.2cm}
      {\bf JAM Collaboration\\
      \vspace*{0.1cm}(Small-$x$ Analysis Group)
        \vspace*{0.2cm}}
        }


\begin{abstract}
We perform a phenomenological study of helicity-dependent parton distribution functions (PDFs) using small-$x$ helicity evolution equations, incorporating for the first time single-inclusive jet production data in polarized proton-proton ($pp$) scattering at parton momentum fractions $x < 0.1$. We also simultaneously include double-longitudinal spin asymmetries in inclusive and semi-inclusive deep-inelastic scattering probing $x < 0.1$.
Employing the polarized small-$x$ pure-glue calculation of $pp\to gX$ for the jet production cross section, we modify the large-$N_c\&N_f$ KPS-CTT evolution equations by setting $N_f = 0$ to replicate the large-$N_c$ (pure-glue) limit, while retaining external quark flavors for the spinor field operators. We find that the $pp$ data have a considerable impact on the helicity PDFs at small $x$, reducing their uncertainties and leading to a total quark and gluon helicity in the proton for $x < 0.1$ of $-0.04 \pm 0.23$. Combining our analysis with the a recent JAM helicity PDF analysis of the world polarized data, which includes $x > 0.1$, we find a total quark and gluon helicity contribution for $x > 10^{-7}$ of between 0.02 and 0.51.
\end{abstract}

\preprint{JLAB-THY-25-4253}

\date{\today}
\maketitle

\section{Introduction}\label{Intro}

Many decades of effort have been devoted to studying the structure of the proton, with one main focus being the so-called proton spin puzzle (see, e.g., Refs.~\cite{Aidala:2012mv, Accardi:2012qut, Leader:2013jra, Aschenauer:2013woa, Aschenauer:2015eha, Boer:2011fh, Proceedings:2020eah, Ji:2020ena, AbdulKhalek:2021gbh}). 
Spin sum rules, such as the Jaffe-Manohar sum rule~\cite{Jaffe:1989jz}, separate the known proton spin of $\tfrac{1}{2}$ (in units of $\hbar$) into distinct terms involving the quark and gluon intrinsic spin ($\Delta\Sigma$, $\Delta G$) and orbital angular momentum ($L_q$, $L_g$),
\begin{equation}\label{spin_sum}
    \frac12 \Delta\Sigma + \Delta G + L_q + L_g = \frac{1}{2}\,,
\end{equation}
where all the terms on the left-hand side depend on $Q^2$.  
The question about how much each piece contributes to the proton spin remains unsettled. 
There also has been considerable discussion about the gauge and renormalization scheme dependence of the individual terms~\cite{Wakamatsu:2010qj, Wakamatsu:2010cb, Hatta:2011zs, Hatta:2011ku, Hatta:2012cs, Leader:2013jra, Ji:2020ena}.
The total quark and gluon (or parton) helicity contributions can be written as integrals of the respective helicity-dependent parton distribution functions (hPDFs),
\begin{subequations}\label{spin_integrals}
\begin{eqnarray} 
    \Delta\Sigma(Q^2) &=& \int\limits_0^1 \dd x\,\sum_q \Delta q^+ (x,Q^2)\equiv \int\limits_0^1 \dd x\, \Delta \Sigma(x,Q^2)\,,
    \\
    \Delta G(Q^2) &=& \int\limits_0^1 \dd x \, \Delta g (x,Q^2) \,,
\end{eqnarray}
\end{subequations}
where $\Delta q^+ \equiv \Delta q + \Delta \bar q$, with $\Delta q$ being the quark hPDF for flavor $q$, $\Delta \bar{q}$ the antiquark hPDF, and $\Delta g$ the gluon hPDF. 
The hPDFs are integrated over the full range of the longitudinal momentum fraction $x$ carried by the corresponding parton.
The goal of this work is to better quantify the quark and gluon hPDFs over the high-energy kinematic region that is associated with $x < 0.1$.
Within collinear factorization, there is a well-defined means to perturbatively evaluate the $Q^2$ dependence of hPDFs through the spin-dependent Dokshitzer--Gribov--Lipatov--Altarelli--Parisi (DGLAP) evolution equations~\cite{Gribov:1972ri, Altarelli:1977zs, Dokshitzer:1977sg} (see, e.g., Chapter~2.4 of Ref.~\cite{Kovchegov:2012mbw}). 
This method allows us to understand the behavior of hPDFs along the $Q^2$-dimension of their $(x,Q^2)$ phase space~\cite{Gluck:2000dy, Leader:2005ci, deFlorian:2009vb, Leader:2010rb, Jimenez-Delgado:2013boa, Ball:2013lla, Nocera:2014gqa, deFlorian:2014yva, Leader:2014uua, Sato:2016tuz, Ethier:2017zbq, DeFlorian:2019xxt, Borsa:2020lsz, Zhou:2022wzm, Cocuzza:2022jye, Borsa:2023tqr, Borsa:2024mss, Hunt-Smith:2024khs, Cruz-Martinez:2025ahf}. 
However, the use of collinear factorization with DGLAP evolution does not predict the small-$x$ behavior of the extracted hPDFs; far below the experimentally measured region, the $x$ dependence is purely an extrapolation.

This lack of predictive power is evidenced by the fact that the fractional error of small-$x$ DGLAP extractions increases rapidly below the $x$-range covered by the fitted data. 
The ability to predict the behavior of hPDFs at small~$x$ is crucial to the proton spin puzzle (recall the $x$-integration limits in Eq.~(\ref{spin_integrals}) go down to zero). 
Since experimental measurements can never probe the proton at $x = 0$, there is an obvious necessity to improve upon the theoretical description of hPDFs at small $x$.

Small-$x$ spin-dependent evolution equations have received renewed interest over the past decade~\cite{Kovchegov:2015pbl, Hatta:2016aoc, Kovchegov:2016zex, Kovchegov:2016weo, Kovchegov:2017jxc, Kovchegov:2017lsr, Kovchegov:2018znm, Kovchegov:2019rrz, Cougoulic:2019aja, Kovchegov:2020hgb, Cougoulic:2020tbc, Chirilli:2021lif, Kovchegov:2021lvz, Cougoulic:2022gbk, Santiago:2024iem, Duan:2024qev, Agrawal:2024xkb, Borden:2024bxa, Kovchegov:2023yzd, Ermolaev:2024aqq, Manley:2024pcl}, building on the recent progress in our understanding of beyond-eikonal corrections to high-energy scattering \cite{Altinoluk:2014oxa,Balitsky:2015qba,Balitsky:2016dgz, Kovchegov:2017lsr, Kovchegov:2018znm, Chirilli:2018kkw, Jalilian-Marian:2018iui, Jalilian-Marian:2019kaf, Altinoluk:2020oyd, Boussarie:2020vzf, Boussarie:2020fpb, Kovchegov:2021iyc, Altinoluk:2021lvu, Kovchegov:2022kyy, Altinoluk:2022jkk, Altinoluk:2023qfr,Altinoluk:2023dww, Li:2023tlw, Altinoluk:2024dba}.
With a focus on small-$x$ helicity evolution, these equations allow one to calculate the $x$ dependence of hPDFs, leading to improved constraints on hPDF uncertainties beyond the $x$-range covered by data~\cite{Adamiak:2021ppq, Adamiak:2023yhz}. 
Generally speaking, the evolution equations are formed from a resummation in the double-logarithmic approximation (DLA) which leads to hPDFs growing as a power law in $1/x$ (outside of the saturation region \cite{Gribov:1984tu, Iancu:2003xm, Weigert:2005us, JalilianMarian:2005jf, Gelis:2010nm, Albacete:2014fwa, Kovchegov:2012mbw, Morreale:2021pnn}), 
\begin{equation}\label{asymptotics}
    \lim_{x\to0} |\Delta \Sigma (x,Q^2)|\,
    \sim\, \lim_{x\to 0}|\Delta g(x,Q^2)|\, 
    \sim\, \biggl(\frac{1}{x}\biggr)^{\!\alpha_h}\,,
\end{equation}
with the exponent $\alpha_h$ representing the helicity intercept. This form was first derived by Bartels, Ermolaev, and Ryskin~(BER) \cite{Bartels:1995iu, Bartels:1996wc} via their use of infrared evolution equations (IREE), and they numerically computed an intercept of $\alpha_h = 3.66\sqrt{\tfrac{\alpha_sN_c}{2\pi}}$ in the pure-gluon case and $\alpha_h = 3.45\sqrt{\tfrac{\alpha_sN_c}{2\pi}}$ for $N_f = 4$ and $N_c = 3$~\cite{Gorshkov:1966ht, Kirschner:1983di, Kirschner:1994rq, Kirschner:1994vc, Griffiths:1999dj}. The intercept for the Kovchegov-Pitonyak-Sievert–Cougoulic-Tarasov-Tawabutr (KPS-CTT) evolution equations was calculated analytically at large-$N_c$ to be $\alpha_h = 3.66074\sqrt{\tfrac{\alpha_sN_c}{2\pi}}$~\cite{Borden:2024jpd}. The expression~\eqref{asymptotics} highlights the theoretical importance of exploring the small-$x$ regime in polarized collisions, as such power law growth has the potential to generate large amounts of spin from small-$x$ partons. 
Furthermore, the currently computed intercepts $\alpha_h$ from BER and KPS-CTT are larger than 1, making the integrals of such hPDFs divergent. The issue of divergent integrals may be resolved via parton saturation~\cite{Gribov:1984tu, Iancu:2003xm, Weigert:2005us, JalilianMarian:2005jf, Gelis:2010nm, Albacete:2014fwa, Kovchegov:2012mbw, Morreale:2021pnn}, which would dampen the hPDF growth \eqref{asymptotics} at very small $x$, or it could be that higher-order corrections in the strong coupling $\alpha_s$ to $\Delta\Sigma(x,Q^2)$ and $\Delta g(x,Q^2)$ result in $\alpha_h < 1$ naturally.

The KPS-CTT small-$x$ helicity evolution equations~\cite{Kovchegov:2015pbl, Hatta:2016aoc, Kovchegov:2016zex, Kovchegov:2016weo, Kovchegov:2017jxc, Kovchegov:2017lsr, Kovchegov:2018znm, Kovchegov:2019rrz, Cougoulic:2019aja, Kovchegov:2020hgb, Cougoulic:2020tbc, Chirilli:2021lif, Kovchegov:2021lvz, Cougoulic:2022gbk}, and the recent global analyses which use them~\cite{Adamiak:2021ppq,Adamiak:2023yhz}, have provided state-of-the-art understanding of small-$x$ spin physics and phenomenology. As an alternative to the BER IREE, the KPS-CTT evolution equations make use of the shock wave or $s$-channel evolution formalism constructed for the case of unpolarized eikonal scattering~\cite{Mueller:1994rr, Mueller:1994jq, Mueller:1995gb, Balitsky:1995ub, Balitsky:1998ya, Kovchegov:1999yj, Kovchegov:1999ua, JalilianMarian:1997dw, JalilianMarian:1997gr, Weigert:2000gi, Iancu:2001ad, Iancu:2000hn, Ferreiro:2001qy}.
In this formalism one finds that $s$-channel gluon cascades seen, for example, in Refs.~\cite{Altinoluk:2014oxa, Balitsky:2015qba, Balitsky:2016dgz, Kovchegov:2017lsr, Kovchegov:2018znm, Chirilli:2018kkw, Jalilian-Marian:2018iui, Jalilian-Marian:2019kaf, Altinoluk:2020oyd, Kovchegov:2021iyc, Altinoluk:2021lvu, Kovchegov:2022kyy, Altinoluk:2022jkk, Altinoluk:2023qfr, Altinoluk:2023dww, Li:2023tlw}, are also present in sub-leading eikonal corrections, with each sub-eikonal order suppressed by an additional power of $x$.
The resummation of diagrams with a power-suppressed sub-eikonal spin-dependent interaction provides the small-$x$ helicity evolution equations~\cite{Kovchegov:2015pbl, Kovchegov:2016zex, Kovchegov:2017lsr, Kovchegov:2018znm, Cougoulic:2022gbk}.

As for the unpolarized small-$x$ evolution equations \cite{Mueller:1994rr, Mueller:1994jq, Mueller:1995gb, Balitsky:1995ub, Balitsky:1998ya, Kovchegov:1999yj, Kovchegov:1999ua, JalilianMarian:1997dw, JalilianMarian:1997gr, Weigert:2000gi, Iancu:2001ad, Iancu:2000hn, Ferreiro:2001qy}, the small-$x$ helicity evolution equations are closed in the large $N_c$ \cite{tHooft:1973alw} and the large-$N_c\&N_f$ \cite{Veneziano:1976wm} limits (with $N_c$ being the number of colors and $N_f$ the number of flavors). These equations evolve polarized dipole amplitudes, which are scattering amplitudes of a color dipole made of lightcone Wilson lines where one or two sub-eikonal operators are inserted to allow for the transfer of spin information~\cite{Kovchegov:2017lsr, Kovchegov:2018znm, Kovchegov:2021iyc, Cougoulic:2022gbk}. 
First developed as the KPS evolution equations~\cite{Kovchegov:2015pbl, Kovchegov:2016zex, Kovchegov:2017lsr}, only \textit{explicitly} spin-dependent operators were considered, which come with a prefactor of $\sigma\,\delta_{\sigma,\sigma'}$ at the sub-eikonal order, with $\sigma$ ($\sigma'$) denoting the incoming (outgoing) quark polarization states.
A large-$N_c$ analysis of these equations resulted in a significantly smaller intercept $\alpha_h$ than that predicted by BER~\cite{Kovchegov:2016weo, Kovchegov:2017jxc}.

The KPS evolution equations have subsequently been updated to include a sub-eikonal operator with the seemingly spin-independent prefactor $\delta_{\sigma,\sigma'}$. However, it has since been shown that this operator can act in a spin-dependent way through a coupling of the orbital angular momentum of the gluon probe to the proton spin~\cite{Cougoulic:2022gbk}. The KPS-CTT evolution equations~\cite{Kovchegov:2015pbl, Kovchegov:2018znm, Cougoulic:2022gbk} include these new spin-dependent contributions. As noted above, numerical and analytical solutions of the KPS-CTT evolution equations at large $N_c$ have been shown to agree with the intercept found by BER up to two decimal places~\cite{Cougoulic:2022gbk, Adamiak:2023okq}, and disagree beyond \cite{Borden:2023ugd}. More recently, in Ref.~\cite{Borden:2024bxa} (see also Ref.~\cite{Chirilli:2021lif}), it was recognized that the quark-to-gluon and gluon-to-quark transition operators also need to be included into small-$x$ helicity evolution beyond large~$N_c$, leading to agreement of the resulting evolution with the three known orders of the spin-dependent DGLAP anomalous dimensions at large $N_c \& N_f$ \cite{Altarelli:1977zs, Dokshitzer:1977sg, Zijlstra:1993sh, Mertig:1995ny, Moch:1999eb, vanNeerven:2000uj, Vermaseren:2005qc, Moch:2014sna, Blumlein:2021ryt, Blumlein:2021lmf, Davies:2022ofz, Blumlein:2022gpp}.  Along with achieving the above theoretical successes, there has also been progress on the numerical and phenomenological~\cite{Kovchegov:2015pbl, Kovchegov:2016zex, Kovchegov:2016weo, Kovchegov:2017lsr, Kovchegov:2018znm, Adamiak:2021ppq, Cougoulic:2022gbk, Adamiak:2023yhz}.

The first analysis~\cite{Adamiak:2021ppq} of polarized data using small-$x$ helicity evolution was conducted on the world polarized inclusive deep-inelastic scattering (DIS) data, and employed the KPS evolution equations. This foundational study was essential to test the ability of small-$x$ theory to describe polarized experimental data, to determine empirically what is sufficiently ``small $x$'' in this formalism, as well as to demonstrate the theoretical control of uncertainties as $x \to 0$. The results showed good agreement with measurements for $x \lesssim 0.1$ and significantly reduced uncertainties, compared to DGLAP-based fits, at $x$ values lower than those covered by the experimental data. However, only the $g_1(x,Q^2)$ structure function could be extracted, since sensitivity to individual quark flavor hPDFs requires additional experimental observables, such as those provided by semi-inclusive deep-inelastic scattering (SIDIS) data.

The latest analysis~\cite{Adamiak:2023yhz} of small-$x$ hPDFs incorporated flavor nonsinglet polarized dipole evolution from Ref.~\cite{Kovchegov:2016zex}, as well as the updated KPS-CTT evolution equations with running coupling in an analysis of DIS and SIDIS data. That analysis was able to resolve quark and antiquark hPDFs for $u$, $d$, and $s$ flavors, and the small-$x$ DIS and SIDIS data were well described. Nevertheless, one of the outcomes~\cite{Adamiak:2023yhz} was a large uncertainty for the small-$x$ extraction of $g_1$, which was attributed to a ``bimodality'' of solutions. Namely, the small-$x$ helicity evolution equations can predict the \textit{growth} of polarized observables at small $x$, but the DIS and SIDIS data were insensitive to the \textit{sign} of the prefactor, allowing the data to be equally well describe with asymptotically positive and negative solutions.

The cause of this is that the small-$x$ behavior of hPDFs and the $g_1$ structure function at fixed $Q^2$ is largely controlled by the gluon-dominated polarized dipole amplitudes $\widetilde{G}$ and $G_2$. 
To reduce the small-$x$ uncertainties in the hPDFs and $g_1$ therefore requires better constraints on $\widetilde{G}$ and $G_2$. 
One approach is to more rigorously model and predict the initial conditions, as was done in Ref.~\cite{Dumitru:2024pcv} and phenomenologically tested in Ref.~\cite{Adamiak:2025mdy}. 
Alternatively, one can incorporate new experimental measurements that are more sensitive to $\widetilde{G}$ and $G_2$ into a phenomenological extraction of the initial conditions. 
We proceed here with the latter method.

Building upon the previous small-$x$ global analysis of polarized DIS and SIDIS data~\cite{Adamiak:2023yhz}, the purpose of this work is to incorporate polarized proton-proton ($pp$) collision data to study the influence on small-$x$ predictions for the hPDFs and the $g_1$ structure function. Proton-proton collisions allow for leading-order (LO) gluon interactions, and as such one expects double-longitudinal spin asymmetries $A_{LL}$ in these reactions to better constrain the gluon-dominated polarized dipole scattering amplitudes $\widetilde{G}$ and $G_2$. Specifically, we include polarized $pp$ data at small $x$ for single-inclusive jet production, $A_{LL}^{\mathrm{jet}}$, at mid-rapidity, using for the numerator of the asymmetry the small-$x$ polarized $pp\to gX$ cross section derived in Ref.~\cite{kovchegov:2024aus}.

The phenomenological implementation of this formula is intended to study the influence of polarized $pp$ data on the small-$x$ extraction of hPDFs. Note that Ref.~\cite{kovchegov:2024aus} derived the gluon production cross section in polarized proton-proton scattering using the small-$x$ or shock wave formalism. The calculation was performed in the gluon-only approximation, with the inclusion of (equally important) quarks left for future work. Therefore, the small-$x$ polarized $pp$ jet production cross section used in this study carries the important caveat that it is not in its final form without including quarks in the $gg\to g$ channel, or explicit quark production. Nevertheless, within the methodology we have defined here, we obtain an unambiguous result that gives us a first indication of the impact of polarized $pp$ data on hPDFs.

As appropriate for the gluon-only parton production channel, we make the choice to use a variation of the large-$N_c$ KPS-CTT evolution equations. A crucial part to our analysis is the need to resolve individual quark flavor hPDFs, a feature that is not provided by the strict large-$N_c$ evolution equations of Ref.~\cite{Cougoulic:2022gbk}. To best approximate a large-$N_c$ analogue in the presence of external quarks, we make use of what we call the ``large-$N_c^{+q}$" evolution equations, which takes the original large-$N_c\&N_f$ KPS-CTT evolution equations from Ref.~\cite{Cougoulic:2022gbk} and sets $N_f = 0$, which eliminates the quark loops but keeps the external quark flavors. 
This permits flavor-dependent initial conditions with flavor-blind evolution. We still include running coupling corrections, which has the effect of reducing the intercept $\alpha_h$ from Eq.~\eqref{asymptotics} and making Eqs.~\eqref{spin_integrals} integrable \cite{Adamiak:2023yhz}. Setting $N_f =0$ also eliminates the corrections found in Ref.~\cite{Borden:2024bxa}, justifying their exclusion from our present analysis.

The rest of this paper is structured as follows. Section~\ref{Formalism} covers updates to our scheme-dependent definition of $\Delta q^{\pm}$ (to align it more closely with the results in the $\overline{\mathrm{MS}}$ scheme), introduces the large-$N_c^{+q}$ helicity evolution equations, and discusses the small-$x$ polarized $pp$ jet production cross section at DLA accuracy. This section also gives details on the numerical implementation of the polarized $pp$ cross section and small-$x$ helicity evolution equations. Section~\ref{Results} outlines our methodology and discusses our results. The definitions of the relevant spin asymmetries for DIS, SIDIS, and polarized $pp$ collisions are provided first, with the results of the global analysis following. We control for the change in evolution equations between Ref.~\cite{Adamiak:2023yhz} and a large-$N_c^{+q}$ analysis of DIS and SIDIS data, and use this to better understand the influence of $pp$ measurements on the extraction of hPDFs and our small-$x$ predictions. The major features we document are the changes to our small-$x$ uncertainty due to the extra constraint from $pp$ data, both on the $g_1^p$ structure function of the proton and the helicity distributions $\Delta\Sigma$ and $\Delta g$. We also mention an important effect the $pp$ data has on the total quark and gluon helicity in the proton. In addition, we perform an Electron-Ion Collider (EIC) impact study as well as carry out a preliminary matching exercise between our small-$x$ $\Delta g$ solutions and the DGLAP-based JAM $\Delta g$ extraction~\cite{Zhou:2022wzm, Hunt-Smith:2024khs}. Finally, in Sec.~\ref{sec:conclusions} we summarize the findings of this analysis, and discuss its implications and possible future extensions.

\section{Theoretical Formalism}\label{Formalism}

\subsection{Helicity PDFs, polarized dipole amplitudes, and small-$x$ helicity evolution}\label{dipole_sec}

The KPS-CTT evolution equations~\cite{Kovchegov:2015pbl, Kovchegov:2018znm, Cougoulic:2022gbk} make use of the light-cone operator treatment (LCOT) framework with polarized dipoles as the degrees of freedom. The important features of these evolution equations are worth mentioning again. They  resum double logarithms of energy through the resummation of parameters $\alpha_s\ln^2(1/x)$ and $\alpha_s\ln(1/x)\ln(Q^2/\Lambda^2)$ (which constitutes the DLA), and the evolution equations close in the large-$N_c$ and/or the large-$N_c\&N_f$ limit. For the purposes of the small-$x$ helicity analysis at hand, the polarized dipole amplitudes themselves \cite{Kovchegov:2018znm, Cougoulic:2022gbk} have already been integrated over the impact parameter and are left dependent on the transverse dipole size $x_{10}$ and the energy of the dipole-proton scattering $zs$, where $z$ is (approximately) the longitudinal momentum fraction of the softest Wilson line in the dipole, and $s$ is the center of mass (CM) energy squared of the projectile-proton scattering. The transverse dipole size is defined as $x_{ij} \equiv |\bm{x}_i - \bm{x}_j|$, with $\bm{x}_{i,j}$ being 2-vectors describing the transverse locations of the Wilson lines in the dipole. As is common in related literature, the transverse dipole size can be denoted as $\bm{x}_{\perp}$ with magnitude $x_{\perp} \equiv x_{ij}$.
We enforce an infrared (IR) cutoff $\Lambda = 1 \, \mathrm{GeV}$ such that $x_{10} < 1/\Lambda$ because the perturbative calculation becomes unreliable at long distances.

In the previous small-$x$ helicity analysis~\cite{Adamiak:2023yhz}, our hPDFs and observables were defined in the ``polarized DIS" scheme.
However, we have implemented some changes to bring our definitions more in line with that of the  $\overline{\mathrm{MS}}$ scheme.
The first of these changes occurs in our definitions of the $C$-even and $C$-odd quark hPDFs at DLA,
\begin{equation}\label{deltaqpm}
    \begin{bmatrix}
         \Delta q^+(x,Q^2) \\
         \Delta q^-(x,Q^2)
     \end{bmatrix}
     = -\frac{N_c}{2\pi^3}\int\limits_{\Lambda^2/s}^1\,\frac{\dd z}{z}\int\limits_{\mathrm{max}\{1/Q^2, 1/zs\}}^{\mathrm{min}\{1/zQ^2,1/\Lambda^2\}}\,\frac{\dd x_{10}^2}{x_{10}^2}\,
     \begin{bmatrix}
         Q_q(x_{10}^2,zs) + 2G_2(x_{10}^2,zs)\\
         -G_q^{\mathrm{NS}}(x_{10}^2,zs)
     \end{bmatrix}\,,
\end{equation}
where $\Delta q^{\pm} \equiv \Delta q \pm \Delta \bar{q}$, and the lower limit of the integral over the (square of the) transverse dipole size has been slightly modified compared to its original form~\cite{Kovchegov:2018znm, Cougoulic:2022gbk}, in accordance with an updated definition in Ref.~\cite{Borden:2024bxa}. This expression allows for a complete agreement to be achieved between KPS-CTT evolution and the $\overline{\mathrm{MS}}$ polarized small-$x$ splitting functions at large $N_c \& N_f$ to the three known loops, up to a residual scheme transformation~\cite{Borden:2024bxa}. The definition for the structure function $g_1(x,Q^2)$ remains the same,
\begin{equation}\label{g1_ex}
    g_1 (x,Q^2) = -\sum_q\frac{N_c\,e_q^2}{4\pi^3}\int\limits_{\Lambda^2/s}^1\,\frac{\dd z}{z}\int\limits_{1/zs}^{\mathrm{min}\{1/zQ^2,1/\Lambda^2\}}\,\frac{\dd x_{10}^2}{x_{10}^2}\,\bigl[Q_q(x_{10}^2,zs) + 2G_2(x_{10}^2,zs)\bigr],
\end{equation}
where $\sum_q$ includes a sum over light quark flavors only ($q = u,d,s$). Consequently, at small~$x$ there is no longer a direct correspondence with the LO collinear factorization expression from polarized DIS,
\begin{equation}\label{g1_relation}
    g_1(x,Q^2) \ne \frac{1}{2}\sum_qe_q^2\,\Delta q^+(x,Q^2)\,.
\end{equation}
Note that the argument $x$ in the hPDF $\Delta q^+(x,Q^2)$ is the parton momentum fraction, while the structure function $g_1$ depends on the Bjorken-$x$ variable. At LO in $\alpha_s$ and in the DLA these are equivalent, but at higher orders the differences need to be taken into account. The DLA form of $\Delta g$ remains consistent with Refs.~\cite{Cougoulic:2022gbk, Adamiak:2023yhz},
\begin{equation}\label{delG}
    \Delta g(x,Q^2) 
    = \frac{2N_c}{\alpha_s\pi^2}\,
    G_2\Bigl(x_{10}^2 = \frac{1}{Q^2},zs = \frac{Q^2}{x}\Bigr).
\end{equation}
The DLA small-$x$ helicity evolution equations resum powers of both $\alpha_s\ln^2(1/x)$ and $\alpha_s\ln(1/x)\ln(Q^2/\Lambda^2)$, and, as such, the definition of the $g_1$ structure function (and the hPDFs) contains information from all orders in those resummation parameters. 
This can be seen through the fact that, from Eq.~\eqref{delG}, $\Delta g \sim G_2$, and $G_2$ also appears in Eq.~\eqref{g1_ex}, which corresponds to the next-to-leading order (NLO) and beyond terms of $g_1$ as derived in the collinear factorization formalism~\cite{Altarelli:1977zs, Dokshitzer:1977sg, Zijlstra:1993sh, Mertig:1995ny, Moch:1999eb, vanNeerven:2000uj, Vermaseren:2005qc, Moch:2014sna, Blumlein:2021ryt, Blumlein:2021lmf, Davies:2022ofz, Blumlein:2022gpp}.

The dipole amplitude $Q_q$ from Ref.~\cite{Cougoulic:2022gbk}, which enters Eqs.~\eqref{deltaqpm} and \eqref{g1_ex}, includes quark flavor dependence $q = u,d,s$, which is necessary to describe the flavor dependent quark spinor field operators. For reasons that will become clear in the next section (a consequence of bringing polarized $pp$ data into this analysis), we must first lay out an approach that can retain this flavor dependence in the large-$N_c$ limit. To combine the large-$N_c$ limit with the  necessary presence of external quarks, we elect to use a modified form of the large-$N_c\&N_f$ evolution equations, that we dub large-$N_c^{+q}$, which takes the  large-$N_c\&N_f$ evolution equations from Ref.~\cite{Cougoulic:2022gbk} and removes the internal quark loops by setting $N_f=0$. This form would best approximate the large-$N_c$ evolution equations with intrinsic flavor dependence. The new evolution equations, both flavor singlet and flavor nonsinglet, are given by
\begin{subequations}\label{eq_LargeNc+q}
\begin{align}
& Q_q (x^2_{10},zs) = Q_q^{(0)}(x^2_{10},zs) + \frac{N_c}{2\pi} \int^{z}_{1/x^2_{10}s} \frac{\dd z'}{z'}   \int_{1/z's}^{x^2_{10}}  \frac{\dd x^2_{21}}{x_{21}^2}  \ \as \!\!\left( \frac{1}{x_{21}^2} \right) \,
\Big[ 2 \, {\widetilde G}(x^2_{21},z's) + 2 \, {\widetilde \Gamma}(x^2_{10},x^2_{21},z's)  
\notag\\
&\hspace*{5cm}+ \; Q_q (x^2_{21},z's) -  \overline{\Gamma}_q (x^2_{10},x^2_{21},z's) + 2 \, \Gamma_2(x^2_{10},x^2_{21},z's) + 2 \, G_2(x^2_{21},z's)   
\Big] 
\notag \\
&\hspace*{3cm}+ \frac{N_c}{4\pi} \int_{\Lambda^2/s}^{z} \frac{\dd z'}{z'}   \int_{1/z's}^{\min \left[ x^2_{10}z/z', 1/\Lambda^2 \right]}  \frac{\dd x^2_{21}}{x_{21}^2} \ \as\!\! \left( \frac{1}{x_{21}^2} \right) \, \left[Q_q (x^2_{21},z's) + 2 \, G_2(x^2_{21},z's) \right],  
\\[0.5cm]
&\overline{\Gamma}_q (x^2_{10},x^2_{21},z's) = Q^{(0)}_q (x^2_{10},z's) + \frac{N_c}{2\pi} \int^{z'}_{1/x^2_{10}s} \frac{\dd z''}{z''}   \int_{1/z''s}^{\min[x^2_{10}, x^2_{21}z'/z'']}  \frac{\dd x^2_{32}}{x_{32}^2}   \ \as\!\! \left( \frac{1}{x_{32}^2} \right) \, \left[ 2\, {\widetilde G} (x^2_{32},z''s)  \right. 
\notag\\
&\hspace*{2.5cm}\left.+ \; 2\, {\widetilde \Gamma} (x^2_{10},x^2_{32},z''s) +  Q_q (x^2_{32},z''s) -  \overline{\Gamma}_q (x^2_{10},x^2_{32},z''s) + 2 \, \Gamma_2(x^2_{10},x^2_{32},z''s) + 2 \, G_2(x^2_{32},z''s) \right] 
\notag \\
&\hspace*{3cm}+ \frac{N_c}{4\pi} \int_{\Lambda^2/s}^{z'} \frac{\dd z''}{z''}   \int_{1/z''s}^{\min \left[ x^2_{21}z'/z'', 1/\Lambda^2 \right] }  \frac{\dd x^2_{32}}{x_{32}^2} \ \as\!\! \left( \frac{1}{x_{32}^2} \right) \, \left[Q_q (x^2_{32},z''s) + 2 \, G_2(x^2_{32},z''s) \right], 
\\[0.5cm]
& {\widetilde G}(x^2_{10},zs) = {\widetilde G}^{(0)}(x^2_{10},zs) + \frac{N_c}{2\pi}\int^{z}_{1/x^2_{10}s}\frac{\dd z'}{z'}\int_{1/z's}^{x^2_{10}} \frac{\dd x^2_{21}}{x^2_{21}} \ \as\!\! \left( \frac{1}{x_{21}^2} \right) \, \left[3 \, {\widetilde G}(x^2_{21},z's) + {\widetilde \Gamma}(x^2_{10},x^2_{21},z's) \right. 
\notag\\[0.25cm]
&\hspace*{4cm}\left.  + \; 2\,G_2(x^2_{21},z's)  +  2\,\Gamma_2(x^2_{10},x^2_{21},z's) \right], 
\\[0.5cm]
& {\widetilde \Gamma} (x^2_{10},x^2_{21},z's) = {\widetilde G}^{(0)}(x^2_{10},z's) + \frac{N_c}{2\pi}\int^{z'}_{1/x^2_{10}s}\frac{\dd z''}{z''}\int_{1/z''s}^{\min[x^2_{10},x^2_{21}z'/z'']} \frac{\dd x^2_{32}}{x^2_{32}} \ \as\!\! \left( \frac{1}{x_{32}^2} \right) \, \left[3 \, {\widetilde G} (x^2_{32},z''s) \right. 
\notag\\[0.25cm]
&\hspace*{4cm}\left. + \; {\widetilde \Gamma}(x^2_{10},x^2_{32},z''s) + 2 \, G_2(x^2_{32},z''s)  +  2\,\Gamma_2(x^2_{10},x^2_{32},z''s) \right], 
\\[0.5cm]
& G_2(x_{10}^2, z s)  =  G_2^{(0)} (x_{10}^2, z s) + \frac{N_c}{\pi} \, \int\limits_{\Lambda^2/s}^z \frac{\dd z'}{z'} \, \int\limits_{\max \left[ x_{10}^2 , \frac{1}{z' s} \right]}^{\min \left[ \frac{z}{z'} x_{10}^2, \frac{1}{\Lambda^2} \right] } \frac{\dd x^2_{21}}{x_{21}^2} \ \as\!\! \left( \frac{1}{x_{21}^2} \right) \, \left[ {\widetilde G} (x^2_{21} , z' s) + 2 \, G_2 (x_{21}^2, z' s)  \right], 
\\[0.5cm]
& \Gamma_2 (x_{10}^2, x_{21}^2, z' s)  =  G_2^{(0)} (x_{10}^2, z' s) + \frac{N_c}{\pi} \!\! \int\limits_{\Lambda^2/s}^{z' \frac{x_{21}^2}{x_{10}^2}} \frac{\dd z''}{z''} \!\!\!\!\!\!\!\!\! \int\limits_{\max \left[ x_{10}^2 , \frac{1}{z'' s} \right]}^{\min \left[  \frac{z'}{z''} x_{21}^2, \frac{1}{\Lambda^2} \right] } \!\! \frac{\dd x^2_{32}}{x_{32}^2} \ \as\!\! \left( \frac{1}{x_{32}^2} \right) \! \left[ {\widetilde G} (x^2_{32} , z'' s) + 2 \, 
G_2(x_{32}^2, z'' s)  \right] , \\[0.5cm]
& G_q^{\mathrm{NS}}(x_{10}^2,zs) = G_q^{\mathrm{NS}\,(0)}(x_{10}^2,zs) + \frac{N_c}{4\pi}\int\limits_{\Lambda^2/s}^z\frac{\dd z'}{z'}\int\limits_{1/z's}^{\mathrm{min}[x_{10}^2z/z',1/\Lambda^2]}\frac{\dd x_{21}^2}{x_{21}^2}\,\alpha_s\!\!\left(\frac{1}{x_{21}^2}\right) G_q^{\mathrm{NS}}(x_{21}^2,z's)\,, \label{NS_evol}
\end{align}
\end{subequations}
where the ``parent" polarized dipole scattering amplitudes $Q_q,\, \widetilde{G},\, G_2$, and $G_q^{\mathrm{NS}}$ are each functions of the transverse dipole size $x_{10}$ and the smallest longitudinal momentum fraction $z$ of the polarized dipole's Wilson lines. 
The ``neighbor" dipole amplitudes $\overline{\Gamma}_q,\, \widetilde{\Gamma},$ and $\Gamma_2$ are so named due to the lifetime cutoff on the subsequent evolution in these dipoles which depends on the size of the adjacent polarized dipole~\cite{Kovchegov:2016zex},
and as such they carry extra transverse-size dependence on $x_{21}$ in addition to the dependence on their own transverse dipole size~$x_{20} \approx x_{10}$.

When comparing directly to the large-$N_c\&N_f$ evolution equations (see Eq.~(155) of Ref.~\cite{Cougoulic:2022gbk}) we see that the $N_f=0$ modification directly affects only the $\widetilde{G}$ and $\widetilde{\Gamma}$ evolution equations by making them independent of the dipole amplitudes $Q_q$ and $\overline{\Gamma}_q$. This means that large-$z$ gluons cannot produce small-$z$ quark loops due to $N_f/N_c$ suppression, which has an effect on the evolution since $\widetilde{G}$ acted as a ``link" that allowed mixing between the $Q_q^{(0)}$ initial conditions and the small-$x$ polarized dipole amplitude $G_2$, and, therefore, the hPDF $\Delta g (x, Q^2)$. The evolution of $\Delta g(x,Q^2)$ still occurs from $\widetilde{G}^{(0)}$ and $G_2^{(0)}$ initial conditions, but the difference between the large-$N_c^{+q}$ and the large-$N_c\&N_f$ evolution equations are worth mentioning. We note that even further updates to the KPS-CTT small-$x$ helicity evolution equations were made in Ref.~\cite{Borden:2024bxa}. In addition to providing the modified versions of the quark helicity PDFs $\Delta q^{\pm}$ seen in Eq.~\eqref{deltaqpm}, Ref.~\cite{Borden:2024bxa} identified two new DLA contributions to the evolution equations regarding quark-to-gluon and gluon-to-quark transition operators. These new contributions introduced a new polarized dipole amplitude $\widetilde{Q}$, which feeds into the evolution of the $\widetilde{G}$ and $\widetilde{\Gamma}$ polarized dipole amplitudes with an $N_f$-enhanced prefactor. Due to this $N_f$ dependence, these updates do not affect our large-$N_c^{+q}$ analysis, as setting $N_f = 0$ decouples $\widetilde{Q}$ from the evolution. 

\subsection{Polarized proton-proton jet production at small $x$}

\begin{figure}[h!]
\begin{centering}
\includegraphics[width=500 pt]{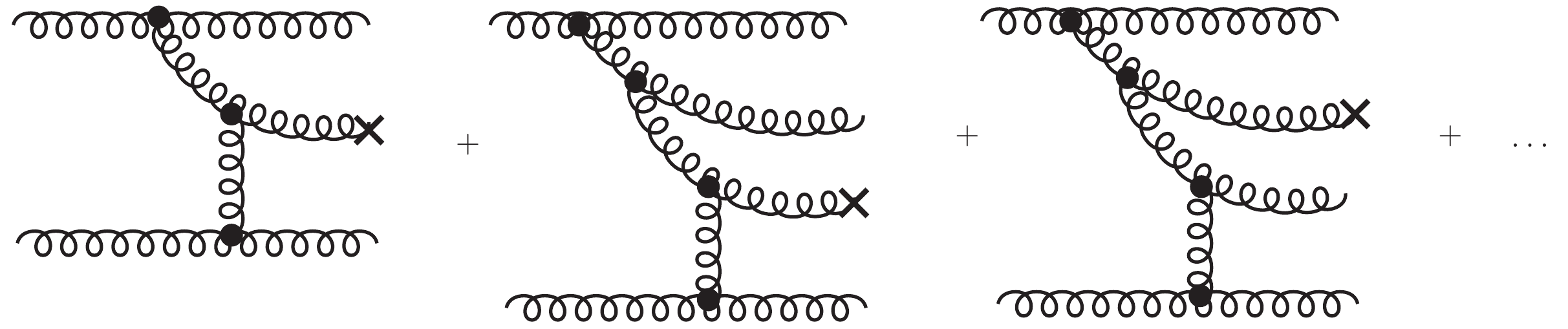}
\caption{Schematic diagram for the process $gg\rightarrow gX$. The black dots indicate the relevant subeikonal-order gluon vertex. 
    \label{gg_to_gluon}
}
\end{centering}
\end{figure}

\begin{figure}[h!]
\begin{centering}
\includegraphics[width=315 pt]{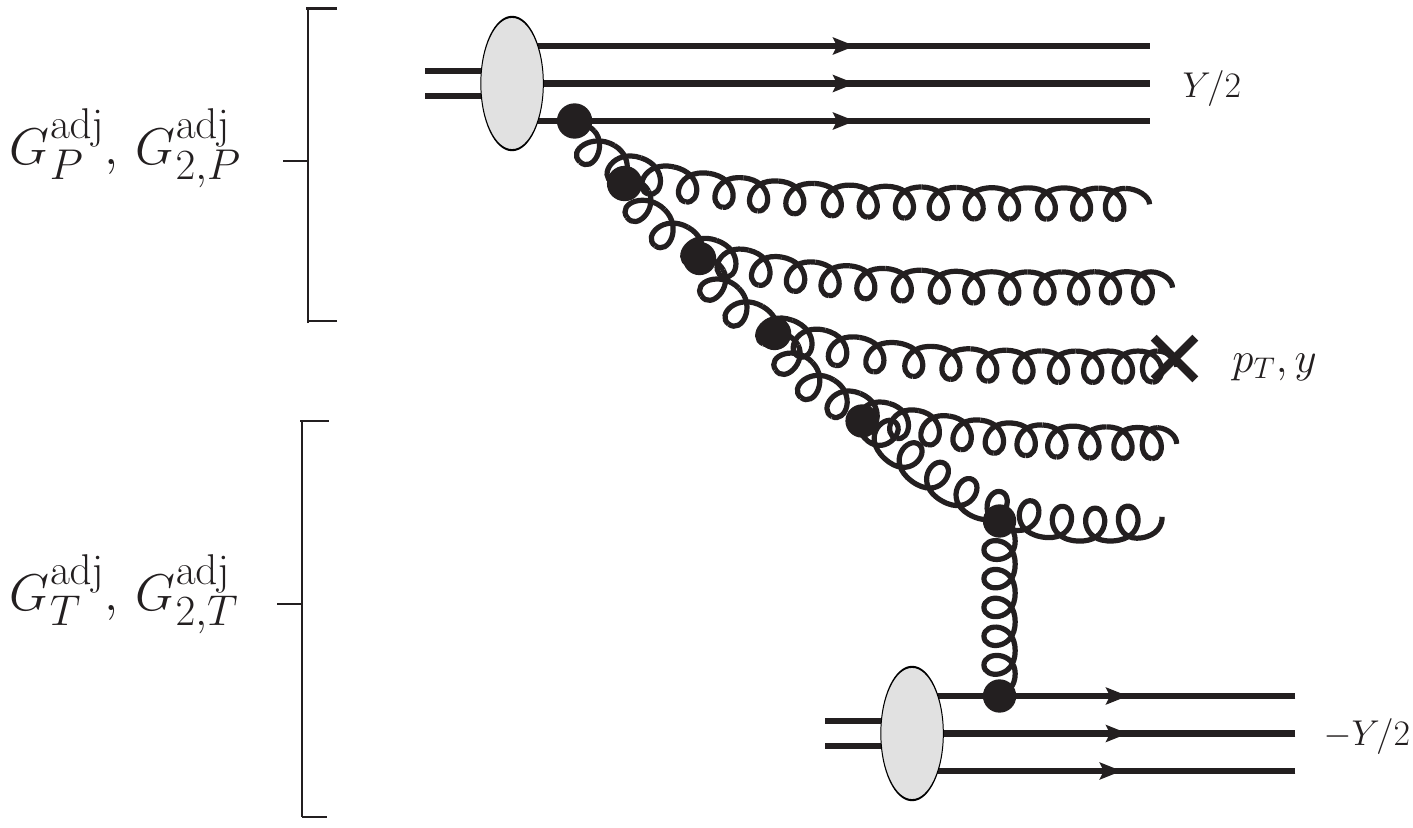}
\caption{Schematic diagram for the process $pp\rightarrow g X$. The projectile and target protons have rapidity $Y/2$ and $-Y/2$, respectively, and the gluon jet is produced at rapidity $y$ in the mid-rapidity region with transverse momentum $p_T$.
    \label{pp_to_gluon}
}
\end{centering}
\end{figure}

The polarized $pp\to gX$ cross section was derived in Ref.~\cite{kovchegov:2024aus} in the small-$x$ helicity formalism, 
\begin{align}\label{KLprod}
    &\frac{\dd\Delta\sigma^{pp\to gX}}{\dd^2\hat{p}_T \dd \hat{y}}
    = \frac{C_F}{\pi^4\alpha_s}\frac{1}{s\,\hat{p}_T^2}\int \dd^2x_{\perp}\,
    e^{-i\hat{\bm{p}}_\perp\cdot\,\bm{x}_\perp}\, 
    \notag \\
    &\hspace{1cm}\times (G_{P}^{\mathrm{adj}}\;G_{2,P}^{\mathrm{adj}})\,\Big(x_{\perp}^2,\sqrt{2}\,p_2^-\,\hat{p}_T\,e^{-\hat{y}}\Big)
    \begin{pmatrix}
        \frac{1}{4}\lvec{\bm{\nabla}}_{\perp}\cdot\rvec{\bm{\nabla}}_{\perp} & \lvec{\bm{\nabla}}_{\perp}^2+\lvec{\bm{\nabla}}_{\perp}\cdot\rvec{\bm{\nabla}}_{\perp} \\
    \rvec{\bm{\nabla}}_{\perp}^2+\lvec{\bm{\nabla}}_{\perp}\cdot\rvec{\bm{\nabla}}_{\perp} & 0
    \end{pmatrix}
    \begin{pmatrix}
        G_{T}^{\mathrm{adj}} \\
        G_{2,T}^{\mathrm{adj}}
    \end{pmatrix}
    \Big(x_{\perp}^2,\sqrt{2}\, p_1^+\, \hat{p}_T\, e^{\hat{y}}\Big)\,,
\end{align}
where $\hat{p}_T = |\hat{\bm{p}}_\perp|$ is the magnitude of the 2-vector transverse momentum of the produced gluon, and $\hat{y} = \frac{1}{2}\ln(\hat{p}^-/\hat{p}^+)$ is its rapidity. 
Note that for DIS and SIDIS we denoted the transverse dipole size as $x_{10}$ because there was only one dipole involved (with Wilson lines positions at 1 and 0). In $pp$ scattering there are two dipole amplitudes (one interacting with the target and one interacting with the projectile\footnote{The choice of language of ``projectile" and ``target" is common in the small-$x$ literature, although not as common when discussing hadronic collisions in collinear factorization.}), so we denote the transverse dipole size more generally as $x_{\perp}$. The large momentum components of the incoming protons are $p_1^+$ and $p_2^-$, which correspond to  rapidities $-Y/2$ and $Y/2$, and the CM energy of the collision is $\sqrt{s}$. The calculation in Ref.~\cite{kovchegov:2024aus} focused on the $gg\to g$ partonic channel, where some typical diagrams in the small-$x$ formalism are shown in Fig.~\ref{gg_to_gluon}.
The polarized dipole amplitudes $G_{P(T)}^{\rm adj}$ and $G_{2,P(T)}^{\rm adj}$ (in the adjoint representation) encode the gluon cascade in the projectile $P$ (target $T$) proton, where one of the gluons is then tagged in the final state --- see Fig.~\ref{pp_to_gluon}.

To the DLA accuracy at which we work, the tagged final-state gluon can be taken to be the detected jet within the framework of the narrow jet approximation (NJA)\cite{Kaufmann:2015hma}.\footnote{The NJA applies to both $k_T$-type and cone-type jets \cite{blazey2000run,Ellis_1993}.} 
In the NJA, the single-inclusive jet production cross section can be written analogously to the usual collinear factorization formula for hadron production, with the typical hadronic fragmentation functions replaced by jet functions $\mathcal{J}(z_j, \frac{R \, p_T}{\mu_F}, \mu_R)$, with jet radius $R$, final state factorization and renormalization scales $\mu_F$ and $\mu_R$, and transverse momentum fractions $z_j$~\cite{Kaufmann:2015hma}.
At the DLA level of accuracy, we use the LO approximation of the jet function, $\mathcal{J} \approx \delta(1-z_j)$. Higher order terms in the jet function may indeed produce logarithms, but these lead to DGLAP evolution of the jet function under $\ln(R)$ resummation rather than small-$x$ evolution. Given that the jet radius of the jet data is not very small,
the inclusion of $\ln(R)$ resummation for jets is left for future work.  Thus, any logarithms in the jet function are beyond the precision of the current analysis.

The polarized $pp\to {\rm jet}\, X$ cross section for small-$R$ jets at small $x$ then reads
\begin{align}\label{num_xperp}
    \frac{\dd\Delta\sigma^{pp\to \mathrm{jet} X}}{\dd^2p_T \dd y} 
    = &\frac{C_F}{\pi^4}\frac{1}{s\,p_T^2}
    \int \dd^2x_\bot\,
    e^{-i\bm{p}_\perp\cdot\, \bm{x}_\perp}\,
    \frac{1}{\alpha_s(1/x_{\perp}^2)}
    \notag \\
    &\times \big( G_{P}^{\mathrm{adj}}\;G_{2,P}^{\mathrm{adj}} \big)\,
    \Big( x_{\perp}^2,\sqrt{s}\,p_T\,e^{-y} \Big)
    \begin{pmatrix}
        \frac{1}{4}\lvec{\bm{\nabla}}_{\perp}\cdot\rvec{\bm{\nabla}}_\perp & \lvec{\bm{\nabla}}_{\perp}^2+\lvec{\bm{\nabla}}_{\perp}\cdot\rvec{\bm{\nabla}}_{\perp} \\
        \rvec{\bm{\nabla}}_{\perp}^2+\lvec{\bm{\nabla}}_{\perp}\cdot\rvec{\bm{\nabla}}_{\perp} & 0
    \end{pmatrix}
    \begin{pmatrix}
        G_{T}^{\mathrm{adj}} \\
        G_{2,T}^{\mathrm{adj}}
    \end{pmatrix}
    \Big( x_{\perp}^2, \sqrt{s}\,p_T\,e^y \Big)\,,
\end{align}
where $\hat{p}_T\to p_T$ and $\hat{y}\to y$ are now the transverse momentum and rapidity of the jet, and in the CM frame we set $p_2^-=p_1^+ = \sqrt{s/2}$. The cross section in Eq.~\eqref{num_xperp} incorporates the running of the strong coupling in a manner inspired by Ref.~\cite{Adamiak:2023yhz}, where we use the standard one-loop expression,
\begin{equation}\label{alpha_s-Q2}
    \alpha_s(Q^2) = \frac{12\,\pi}{11N_c - 2N_f}\frac{1}{\ln\!\big(Q^2/\Lambda_{\rm QCD}^2\big)} \,
\end{equation}
replacing $Q^2 \to 1/x_{\perp}^2$ when computing $\alpha_s$ in \eq{num_xperp}. The QCD confinement scale $\Lambda_{\mathrm{QCD}}$ is taken to be $241~\mathrm{MeV}$ such that $\alpha_s(M_Z^2) = 0.1176$ for $N_f=3$, with $M_z$ the mass of the $Z$ boson~\cite{Albacete:2009fh} (note that our IR cutoff is such that $\Lambda > \Lambda_{\mathrm{QCD}}$). For the $Q^2$ range we consider in this analysis, we compute the running coupling with $N_c = 3$ and $N_f = 3$.

In the large-$N_c$ limit we can make use of the relations $G^{\mathrm{adj}} = 4 \widetilde{G}$ and $G_2^{\mathrm{adj}} = 2G_2$ \cite{Kovchegov:2018znm, Cougoulic:2022gbk} that connect polarized dipole amplitudes in the adjoint and fundamental representations. 
Making this substitution and expanding the matrix multiplication, we recover the quadratic dependence on the gluon dipoles of the polarized $pp$ jet production cross section:
\begin{align}
    (G_{P}^{\mathrm{adj}}\;&G_{2,P}^{\mathrm{adj}})\,
    \begin{pmatrix}
        \frac{1}{4}\lvec{\bm{\nabla}}_{\perp}\cdot\rvec{\bm{\nabla}}_{\perp} & \lvec{\bm{\nabla}}_{\perp}^2+\lvec{\bm{\nabla}}_{\perp}\cdot\rvec{\bm{\nabla}}_{\perp} \\
        \rvec{\bm{\nabla}}_{\perp}^2+\lvec{\bm{\nabla}}_{\perp}\cdot\rvec{\bm{\nabla}}_{\perp} & 0
    \end{pmatrix}
    \begin{pmatrix}
        G_{T}^{\mathrm{adj}} \\
        G_{2T}^{\mathrm{adj}}
    \end{pmatrix} 
    \notag \\
    & = 4\Bigl[2\,G_{2,P}\bm{\nabla}_{\perp}^2\widetilde{G}_T + 2\,(\bm{\nabla}_{\perp}^2\widetilde{G}_P)G_{2,T} + \frac{\partial}{\partial x_{\perp}}\widetilde{G}_P\frac{\partial}{\partial x_{\perp}}\widetilde{G}_T + 2\,\frac{\partial}{\partial x_{\perp}}G_{2,P}\frac{\partial}{\partial x_{\perp}}\widetilde{G}_T + 2\,\frac{\partial}{\partial x_{\perp}}\widetilde{G}_P\frac{\partial}{\partial x_{\perp}}G_{2,T}\Bigr] \,,
\end{align}
where we have suppressed the arguments of the dipole amplitudes for brevity and have made use of the fact that the impact-parameter integrated dipole amplitudes are azimuthally symmetric for the target proton. The quadratic dependence on the gluon dipoles is an essential feature of the $pp$ cross section for the gluon production at mid-rapidity. The DIS and SIDIS cross sections are strictly linear in the gluonic polarized dipole amplitudes, which is to be expected since gluons can only be produced from the (single) nucleon target. This is contrasted with $pp$ collisions, where both the proton projectile and the proton target contain gluons, resulting in a quadratic dependence on the dipole amplitudes. We then integrate over the angles of $\bm{x}_\perp$ to obtain our final expression for the differential cross section, \\
\begin{align}\label{jet_numerator}
    &\frac{\dd\Delta\sigma^{pp\to \mathrm{jet} X}}{\dd^2p_T \dd y} 
    = \frac{8\,C_F}{\pi^3}\frac{1}{s\,p_T^2}\int\limits_0^{\infty} \dd x_{\perp} \frac{1}{\alpha_s(1/x_\perp^2)}\, x_{\perp} J_0(p_T\, x_{\perp}) 
    \notag \\
    &\hspace{1cm} \times  \Bigl[2\,G_{2,P}\bm{\nabla}_{\perp}^2\widetilde{G}_T + 2\,(\bm{\nabla}_{\perp}^2\widetilde{G}_P)G_{2,T} + \frac{\partial}{\partial x_{\perp}}\widetilde{G}_P\frac{\partial}{\partial x_{\perp}}\widetilde{G}_T + 2\,\frac{\partial}{\partial x_{\perp}}G_{2,P}\frac{\partial}{\partial x_{\perp}}\widetilde{G}_T + 2\,\frac{\partial}{\partial x_{\perp}}\widetilde{G}_P\frac{\partial}{\partial x_{\perp}}G_{2,T}\Bigr]\,.
\end{align}
We will now discuss the numerical implementation of this formula and that of the evolution equations~\eqref{eq_LargeNc+q}. 

\subsection{Numerical implementation}

Following the numerical approach of Ref.~\cite{Adamiak:2023yhz}, we recast the evolution equations into the logarithmic variables defined below, simplify them using recursion relations, discretize them into an evenly spaced $\eta$-$s_{10}$ grid ($\Delta \eta = \Delta s_{10} \equiv \Delta$) and compute the remaining integrals via left-hand Riemann sums. In this format we write the polarized dipole amplitudes' arguments as $i,\,j$ (and $k$ for neighbor dipoles) which are to be understood as integers $i,j, (k) =0,1,2, ...$ and implicitly scaled by the grid spacing such that, e.g., $G[i,(k),j]\equiv G(i\Delta, (k\Delta), j\Delta)$. The discretized evolution equations are largely the same as those used in the large-$N_c\&N_f$ analysis \cite{Adamiak:2023yhz}, but the large-$N_c^{+q}$ evolution equations~\eqref{eq_LargeNc+q} differ in the evolution kernels of $\widetilde{G}$ and $\widetilde{\Gamma}$. As previous papers have shown \cite{Kovchegov:2016weo, Kovchegov:2020hgb, Cougoulic:2022gbk, Adamiak:2023okq, Adamiak:2023yhz}, the evolutions equations are much simpler in the logarithmic variables 
\begin{equation}\label{varchange}
    \eta^{(n)} \equiv \sqrt{\frac{N_c}{2\pi}}\, \ln\!\bigg(\frac{z^{(n)}s}{\Lambda^2}\bigg),\qquad 
    s_{ij} \equiv \sqrt{\frac{N_c}{2\pi}}\, \ln \!\bigg(\frac{1}{x_{ij}^2\Lambda^2}\bigg),
\end{equation}
with $\eta^{(n)} = \eta, \eta', \eta''$ and $z^{(n)} = z, z', z''$, where the primed arguments belong to the daughter dipoles that are being integrated in Eqs.~\eqref{eq_LargeNc+q}. We note that the evolution equations~\eqref{eq_LargeNc+q} are only valid for small $x$, yet there is no explicit constraint enforcing that condition. Therefore, we introduce a small-$x$ cutoff $x_0$ and demand that the evolution of the polarized dipole amplitudes only starts at $x<x_0$.  In terms of the logarithmic variables~\eqref{varchange}, this constraint corresponds to $\eta - s_{10} > \sqrt{\tfrac{N_c}{2\pi}}\ln(1/x_0) \equiv y_0$. This extends to the daughter dipoles, where we also impose $\eta'-s_{21}>y_0$ and $\eta'' - s_{32} > y_0$.

For $x>x_0$, the evolution equations simply output their (comparably small) initial conditions. Additionally, the perturbative calculation breaks down for dipole sizes larger than $1/\Lambda$, which roughly characterizes the size of the proton.  This is enforced by the condition $s_{ij} > 0$. There also must be a lower limit for the transverse size of the dipole, defined from the smallest length-scale in the problem, $1/zs$. Incorporating each of these constraints into Eqs.~\eqref{eq_LargeNc+q}, and then carrying out the discretization procedure described above, the discretized form of the large-$N_c^{+q}$ evolution equations can be written as
\begin{subequations}\label{DiscreteEvol}
    \begin{align}
        Q_q[i,j] & = Q_q[i,j-1] + Q_q^{(0)}[i,j] -Q_q^{(0)}[i,j-1] \label{DiscreteEvol_Q} \notag\\
        &\;\;\;\;\;+ \Delta^2\sum_{i'=i}^{j-2-j_0}\,\alpha_s[i']\Bigl[\frac{3}{2}Q_q[i',j-1]+2\widetilde{G}[i',j-1]+2\widetilde{\Gamma}[i,i',j-1] 
        \notag \\
        &\hspace{4cm}-\overline{\Gamma}_q[i,i',j-1]+3G_2[i',j-1]+2\Gamma_2[i,i',j-1]\Bigr] \notag  \\
        &\;\;\;\;\;+ \frac{1}{2}\Delta^2\sum_{j'=j-1-i}^{j-2}\,\alpha_s[i+j'-j+1]\Bigl[Q_q[i+j'-j+1,j'] + 2G_2[i+j'-j+1,j']\Bigr] \, ,  
        \\[0.5cm]
        \overline{\Gamma}_q[i,k,j] &= \overline{\Gamma}_q[i,k-1,j-1]+Q_q^{(0)}[i,j] - Q_q^{(0)}[i,j-1] \label{DiscreteEvol_Gmb} \notag\\
        &\;\;\;\;\;+ \Delta^2\sum_{i'=k-1}^{j-2-j_0}\,\alpha_s[i']\Bigl[\frac{3}{2}Q_q[i',j-1]+2\widetilde{G}[i',j-1]+2\widetilde{\Gamma}[i,i',j-1] \notag \\
        &\hspace{4cm}-\overline{\Gamma}_q[i,i',j-1]+3G_2[i',j-1]+2\Gamma_2[i,i',j-1]\Bigr] \, ,  
        \\[0.5cm]
        \notag\\
        \widetilde{G}[i,j] & = \widetilde{G}[i,j-1] + \widetilde{G}^{(0)}[i,j] - \widetilde{G}^{(0)}[i,j-1] \label{DiscreteEvol_Gt} 
        \notag\\ 
        &\;\;\;\;\;+\Delta^2\sum_{i'=i}^{j-2-j_0}\,\alpha_s[i']\Bigl[3\widetilde{G}[i',j-1]+\widetilde{\Gamma}[i,i',j-1] +2G_2[i',j-1]+2\Gamma_2[i,i',j-1]\Bigr],
        \\[0.5cm]
        \widetilde{\Gamma}[i,k,j] &= \widetilde{\Gamma}[i,k-1,j-1]+ \widetilde{G}^{(0)}[i,j] - \widetilde{G}^{(0)}[i,j-1] \label{DiscreteEvol_Gmt} 
        \notag\\
        &\;\;\;\;\;+ \Delta^2\sum_{i'=k-1}^{j-2-j_0}\,\alpha_s[i']\Bigl[3\widetilde{G}[i',j-1]+\widetilde{\Gamma}[i,i',j-1] +2G_2[i',j-1]+2\Gamma_2[i,i',j-1]\Bigr],  
        \\[0.5cm]
        G_2[i,j] &= G_2[i,j-1] + G_2^{(0)}[i,j]-G_2^{(0)}[i,j-1]  \label{DiscreteEvol_G2} \notag\\
        &\;\;\;\;\;+2\Delta^2\sum_{j'=j-1-i}^{j-2}\,\alpha_s[i+j'-j+1]\Bigl[\widetilde{G}[i+j'-j+1,j']+2G_2[i+j'-j+1,j']\Bigr] \, ,   
        \\[0.5cm]
        \Gamma_2[i,k,j] &= \Gamma_2[i,k-1,j-1] + G_2^{(0)}[i,j] - G_2^{(0)}[i,j-1]   \, , \label{DiscreteEvol_Gm2}
        \\[0.5cm]
        G_q^\mathrm{NS}[i,j] & = G_q^\mathrm{NS}[i,j-1]+G_q^{\mathrm{NS} \,(0)}[i,j] - G_q^{\mathrm{NS} \,(0)}[i,j-1] \notag\\
        &\;\;\;\;\;+ \frac{1}{2}\Delta^2\Biggl[\sum\limits_{i'=i}^{j-2-j_0}\alpha_s[i'] \, G_q^\mathrm{NS}[i',j-1]+\sum\limits_{j'=j-1-i}^{j-2} \as[i-j+1+j'] \, G_q^\mathrm{NS}[i-j+1+j',j']\Biggr] . 
\end{align}
\end{subequations}
where $j_0 \equiv \lceil y_0/\Delta \rceil$.  As discussed in Sec.~\ref{dipole_sec}, $\widetilde{G}$ and $\widetilde{\Gamma}$ are now notably missing any dependence on $\Gamma_q$ or $Q_q$ as compared to the analogous formulas that appeared in Eqs.~(17) and (23) of Ref.~\cite{Adamiak:2023yhz}.

One other numerical difference between this and our previous analysis is the need to compute the derivatives of the polarized dipole amplitudes in Eq.~\eqref{jet_numerator}. For this we implement the $\mathcal{O}(\Delta^3)$ backward finite difference approximation for the first and second derivatives: 
\begin{subequations}
\begin{align}
    f'(x)&\approx \frac{1}{\Delta} \Big( \tfrac{11}{6}f(x) - 3f(x-\Delta) + \tfrac{3}{2}f(x-2\Delta) - \tfrac{1}{3}f(x-3\Delta)\Big) \,,
    \\
    f''(x)&\approx \frac{1}{\Delta^2} \Big( \tfrac{35}{12}f(x) - \tfrac{26}{3}f(x-\Delta) + \tfrac{19}{2}f(x-2\Delta) - \tfrac{14}{3}f(x-3\Delta) + \tfrac{11}{12}f(x-4\Delta) \Big)\,.
\end{align}
\end{subequations}
Note that numerical integration is more accurate than numerical differentiation; our previous analysis, having only computed numerical integrals, had a numerical uncertainty well below 1\%. 
This analysis, with additionally needing to compute numerical derivatives given the same computational resources, has a numerical uncertainty of approximately 2.5\%.  Including higher-order finite difference approximations showed negligible improvement beyond this.

Although the evolution equations~\eqref{DiscreteEvol} now encode some important physical properties and constraints on the polarized dipole amplitudes, those features still must be carried over to the hPDFs, polarized structure functions, and the polarized $pp$ jet production cross section. We thus impose the constraints $\eta - s_{10} > y_0$ and $s_{10} > 0$ to account for the small-$x$ cutoff and the upper limit for the transverse dipole size, respectively. 
The lower limit for the transverse dipole size, however, requires more careful consideration.

When dealing with (SI)DIS processes, the dipole amplitudes have a transparent physical correspondence to actual $q \bar{q}$ dipoles produced by quantum fluctuations of the virtual photon, so when the cross section is factorized into polarized dipole amplitudes $Q_q(x_{10}^2, z s)$ or $G_2(x_{10}^2,zs)$, there is one transverse dipole size that has sensitivity to the shortest length scale $x_{10}^2 \geq 1 / z s$ as described for the evolution equations. 
For jet production in polarized $pp$ collisions, however, the correspondence is less obvious since there are the two distinct dipole amplitudes with differing longitudinal momentum fractions $z^{(P,T)}$ with respect to the projectile moving along the $-z$-axis or the target moving along the $+z$-axis. Additionally, the transverse size is Fourier-conjugate to the jet $p_T$ such that $x_\bot \sim \ord{1 / p_T}$. The distinct relevant longitudinal momentum fractions are determined by both the transverse momentum $p_T$ and rapidity $y$ of the produced jet \cite{kovchegov:2024aus}, 
\begin{align}
    z^{(P)} \equiv \frac{p_T}{\sqrt{s}} e^{y}\,
    , \qquad
    z^{(T)} \equiv \frac{p_T}{\sqrt{s}} e^{- y}
    \: .
\end{align}
The short-distance cutoff for $pp$ collisions then results in $1/x_{\perp}^2 < \mathrm{min}\{z^{(T)} s,\,z^{(P)} s\}$. 
These assumptions and our change of variables~\eqref{varchange} must now be consistently enforced onto our polarized $pp$ cross section.

From the dipole variables $(x_\bot, z^{(P/T)}s)$ one may similarly define logarithmic variables $(s_{10}, \eta^{(P/T)})$ by replacing $x_{10}\to x_{\perp}$ and $z\to z^{(P/T)}$ in Eq.~\eqref{varchange}. The change of variables of course also extends to the derivatives with respect~to~$x_{\perp}$,
\begin{subequations}
\begin{align}
    \frac{\partial}{\partial x_{\perp}} 
    &= -\Lambda \sqrt{\frac{2N_c}{\pi}} \exp\!\left(\sqrt{\frac{\pi}{2N_c}} s_{10}\right) \frac{\partial}{\partial s_{10}}\,,
    \\
    \bm{\nabla}_{\perp}^2 
    &= \frac{1}{x_{\perp}}\frac{\partial}{\partial x_{\perp}} + \left( \frac{\partial}{\partial x_{\perp}} \right)^2
    = \Lambda^2\frac{2N_c}{\pi} 
    \exp\!\left(\sqrt{\frac{2\pi}{N_c}} s_{10}\right) 
    \left( \frac{\partial}{\partial s_{10}} \right)^2 \,,
\end{align}
\end{subequations}
and the running coupling \eqref{alpha_s-Q2},
\begin{equation}\label{alpha_s-s10}
    \alpha_s(s_{10}) = \sqrt{\frac{N_c}{2\,\pi}}\frac{12\,\pi}{11 N_c-2 N_f}\frac{1}{(s_{10}+s_0)},\qquad s_0 \equiv \sqrt{\frac{N_c}{2\pi}}\ln\frac{\Lambda^2}{\Lambda_{\rm QCD}^2}\,,
\end{equation}
where we avoid the Landau pole at $s_{10} = -s_0$ since $s_{10},\,s_0 > 0$. 
Combining all these modifications into Eq.~\eqref{jet_numerator}, we can now write 
\begin{align}\label{pJet_GA_Bessel}
    \frac{\dd\Delta\sigma^{pp\to \mathrm{jet} X}}{\dd^2 p_T\,\dd y} 
    = &\frac{16\,C_F}{\pi^3}\frac{1}{p_T^2\,s}\sqrt{\frac{N_c}{2\pi}}\,\int\limits_{0}^{\mathrm{min}\{\eta^{(P)},\eta^{(T)}\}}
    \dd s_{10}\,
    \frac{1}{\alpha_s(s_{10})}\, J_0\biggl(\frac{p_T}{\Lambda}
    \exp\Big(-\sqrt{\frac{\pi}{2 N_c}} s_{10}\Big)\biggr)
    \notag \\
    &\times \Bigl\{2\,G_2(s_{10},\eta^{(T)})\widetilde{G}^{(\prime\prime)}(s_{10},\eta^{(P)}) + 2\,\widetilde{G}^{(\prime\prime)}(s_{10},\eta^{(T)})G_2(s_{10},\eta^{(P)})+\widetilde{G}^{(\prime)}(s_{10},\eta^{(T)})\widetilde{G}^{(\prime)}(s_{10},\eta^{(P)}) \notag \\
    &\hspace{15pt}+2\,G_2^{(\prime)}(s_{10},\eta^{(T)})\widetilde{G}^{(\prime)}(s_{10},\eta^{(P)})+ 2\,\widetilde{G}^{(\prime)}(s_{10},\eta^{(T)})G_2^{(\prime)}(s_{10},\eta^{(P)})\Bigr\}
    \bigg|_{\eta^{(P,T)}=\sqrt{\frac{N_c}{2\pi}}\left(\ln\frac{p_T\sqrt{s}}{\Lambda^2}\pm y\right)}\,,
\end{align}
where $G^{(\prime),(\prime\prime)}$ refer to the first and second derivatives of the dipole amplitudes with respect to $s_{10}$. We point out that the upper limit of integration in Eq.~\eqref{pJet_GA_Bessel} is due to the fact that $s_{10}<\eta$ in the DLA. In arriving at \eq{pJet_GA_Bessel} we have also imposed the $s_{10} >0$ cutoff on the integral, restricting the dipole sizes to $x_\perp < 1/\Lambda$. These sharp cutoffs on the integral, though, cause oscillation artifacts from the $J_0$ Bessel function that are not intended features of small-$x$ evolution in the DLA. Therefore, to avoid these oscillation artifacts, we impose $\sqrt{\tfrac{2N_c}{\pi}}\ln(p_T/\Lambda)$ as a lower limit on the $s_{10}$ integration and set $J_0 = 1$. The final form for the cross section that we use in our global analysis is then
\begin{align}\label{pJet_GA}
    \frac{\dd\Delta\sigma^{pp\to \mathrm{jet} X}}{\dd^2 p_T\, \dd y}
    = &\frac{16\,C_F}{\pi^3}\frac{1}{p_T^2\,s}\sqrt{\frac{N_c}{2\pi}}\,\int\limits_{\sqrt{\frac{2N_c}{\pi}}\ln(p_T/\Lambda)}^{\mathrm{min}\{\eta^{(P)},\eta^{(T)}\}}\dd s_{10}\,\frac{1}{\alpha_s(s_{10})} \notag \\
    &\times \Bigl\{2\,G_2(s_{10},\eta^{(T)})\widetilde{G}^{(\prime\prime)}(s_{10},\eta^{(P)}) + 2\,\widetilde{G}^{(\prime\prime)}(s_{10},\eta^{(T)})G_2(s_{10},\eta^{(P)})+\widetilde{G}^{(\prime)}(s_{10},\eta^{(T)})\widetilde{G}^{(\prime)}(s_{10},\eta^{(P)}) \notag \\
    &\hspace{15pt}+2\,G_2^{(\prime)}(s_{10},\eta^{(T)})\widetilde{G}^{(\prime)}(s_{10},\eta^{(P)})+ 2\,\widetilde{G}^{(\prime)}(s_{10},\eta^{(T)})G_2^{(\prime)}(s_{10},\eta^{(P)})\Bigr\} \bigg|_{\eta^{(P,T)}=\sqrt{\frac{N_c}{2\pi}}\left(\ln\frac{p_T\sqrt{s}}{\Lambda^2}\pm y\right)}\,.
\end{align}

We can rewrite the hPDFs and polarized structure functions to include these physical assumptions and changes of variables. With the modified integration limits for the ${\rm \overline{MS}}$-consistent definitions in Eq.~\eqref{deltaqpm}, The hPDFs are given by
\begin{subequations}\label{hPDFs_log}
\begin{align}
    \Delta q^+ (x, Q^2) &= -\frac{1}{\pi^2} \int\limits_0^{\sqrt{\frac{N_c}{2\pi}}\ln\frac{Q^2}{\Lambda^2}}\,\dd s_{10}\,\int\limits_{s_{10}+ y_0}^{s_{10}+\sqrt{\frac{N_c}{2\pi}}\ln\frac{1}{x}} \dd\eta \, \Bigl[ Q_q(s_{10}, \eta) + 2G_2(s_{10},\eta) \Bigr]\,, 
    \\
    \Delta q^- (x, Q^2) &= \frac{1}{\pi^2} \int\limits_0^{\sqrt{\frac{N_c}{2\pi}}\ln\frac{Q^2}{\Lambda^2}}\,\dd s_{10}\,\int\limits_{s_{10}+y_0}^{s_{10}+\sqrt{\frac{N_c}{2\pi}}\ln\frac{1}{x}} \dd \eta \, G_q^{\mathrm{NS}}(s_{10},\eta)\,, 
    \\[0.1cm]
    \Delta g(x,Q^2) &= \frac{2N_c}{\alpha_s(Q^2)\pi^2}G_2(s_{10}, \eta)\Bigg|_{\eta = \sqrt{\frac{N_c}{2\pi}}\ln\frac{Q^2}{x\Lambda^2}}^{s_{10} = \sqrt{\frac{N_c}{2\pi}}\ln\frac{Q^2}{\Lambda^2}} \,,
\end{align}
\end{subequations}
and the polarized DIS and SIDIS structure functions, respectively, are
\begin{subequations}\label{g1_log}
\begin{align}
    g_1(x,Q^2) &= -\frac{1}{\pi^2}\sum_q e_q^2 \int\limits_{y_0}^{\sqrt{\frac{N_c}{2\pi}}\ln\frac{Q^2}{x\Lambda^2}} \dd\eta\,\int\limits_{\mathrm{max}\{0, \eta - \sqrt{\frac{N_c}{2\pi}}\ln\frac{1}{x}\}}^{\eta - y_0} \dd s_{10} \, \Bigl[Q_q(s_{10}, \eta) + 2G_2(s_{10},\eta)\Bigr] \,,
    \\[0.1cm]
    g_1^h(x, z_h, Q^2)&= -\frac{1}{4\pi^2} \sum_q e_q^2 \, D_q^h(z_h,Q^2) 
    \int\limits_{y_0}^{\sqrt{\frac{N_c}{2\pi}}\ln\frac{Q^2}{x\Lambda^2}} \dd\eta\,\int\limits_{\mathrm{max}\{0, \eta - \sqrt{\frac{N_c}{2\pi}}\ln\frac{1}{x}\}}^{\eta - y_0} \dd s_{10} \, \Bigl[Q_q(s_{10}, \eta) + 2G_2(s_{10},\eta) \pm G_q^{\mathrm{NS}}(s_{10},\eta)\Bigr],
    \label{g1h}
\end{align}
\end{subequations}
where $D_q^h(z_h,Q^2)$ is the unpolarized fragmentation function expressed in terms of the fraction $z_h$ of the fragmenting parton's light-cone momentum carried by the observed hadron $h$. Note that the flavor nonsinglet polarized dipole amplitude $G_q^{\mathrm{NS}}$ enters in the polarized SIDIS structure function; the relative plus or minus sign is due to whether the tagged hadron came from a quark or antiquark, since $\Delta q = \tfrac{1}{2}(\Delta q^+ + \Delta q^-)$ and $\Delta \bar{q} = \tfrac{1}{2}(\Delta q^+ - \Delta q^-)$.

Our choice of initial conditions are the same as those in Ref.~\cite{Adamiak:2023yhz}, where a linear-expansion ansatz (in terms of the logarithmic variables) is employed and the initial conditions are parameterized as
\begin{subequations}\label{Dipole_ICs}
\begin{align}
    Q_q^{(0)}(s_{10},\eta) &= a_q\,\eta + b_q\,s_{10} + c_q\,, \\
    \widetilde{G}^{(0)}(s_{10},\eta) &= \widetilde{a}\,\eta + \widetilde{b}\,s_{10}+\widetilde{c}\,, \\
    G_2^{(0)}(s_{10},\eta) &= a_2\,\eta+b_2\,s_{10}+c_2\,, \\
    G_q^{\mathrm{NS},(0)}(s_{10},\eta) &= a_q^{\mathrm{NS}}\,\eta + b_q^{\mathrm{NS}}\,s_{10} + c_q^{\mathrm{NS}},
\end{align}
\end{subequations}
with the parameters $a,b,c$ extracted from fitting the experimental data. The evolution equations, hPDFs, and polarized structure functions all depend only linearly on the polarized dipole amplitudes, and thus the hPDFs and polarized structure functions can be expressed as an arbitrary linear combination of ``basis" dipole amplitudes.\footnote{As defined in Ref.~\cite{Adamiak:2023yhz}, the ``basis" dipole amplitudes are an evolved set of polarized dipole amplitudes constructed solely out of only one initial condition parameter. That is, we can construct a ``basis" hPDF out of the polarized dipole amplitudes that were evolved from an initial condition where $a_u = 1$ and $b_u = c_u = a_d = b_d = \cdots = 0$.}  
However, the polarized proton-proton jet production cross section cannot be expressed as a linear combination of basis functions in the same way since its dependence on the polarized dipole amplitudes is nonlinear.

Lastly, we mention that using large-$N_c^{+q}$ evolution equations retains an important behavior observed in Ref.~\cite{Adamiak:2023yhz}, namely that basis hPDFs constructed from $G_2^{(0)}$ and $\widetilde{G}^{(0)}$ initial conditions have the two largest contributions at small $x$.  Since these are the two polarized dipole amplitudes for which the $pp$ jet production cross section \eqref{pJet_GA} is directly sensitive, the constraints on parameters $\widetilde{a},\, \widetilde{b},\, \widetilde{c},\, a_2,\, b_2,\, c_2$ from polarized $pp$ data are especially impactful on the small-$x$ hPDF extractions discussed in the following section.

\section{Global QCD analysis} \label{Results}

In this section we present the results of our global analysis.
First, in Sec.~\ref{ssec:methodology} we focus on the methodology of our fit, and provide the relevant theoretical formulas for the experimental observables and a description of the kinematic cuts applied to the data.
In Sec.~\ref{ssec:data_vs_thy}, we assess the quality of our fit using the $\chi^2$ statistics.
We enumerate all of the experimental datasets included in our analysis and present tables of the $\chi^2$, for each dataset individually as well as in aggregate, obtained when using our formalism to describe the measurements. We also explicitly plot a comparison of data versus theory for $A_{LL}^{\rm jet}$.
Successfully completing a fit to the data gives posterior distributions for the undetermined parameters of the initial conditions. 
These can then be used to compute the proton structure function $g_1^p$ and the hPDFs, which we present in Sec.~\ref{ssec:effectofpp}.
Importantly, we discuss the impact of including the polarized $pp$ data on the helicity distributions and on our estimates of the amount of proton spin arising from the helicities of small-$x$ quarks and gluons, where we compare our current results with the previous polarized DIS+SIDIS only small-$x$ analysis in Ref.~\cite{Adamiak:2023yhz}. 
Lastly, in Sec.~\ref{ssec:matching}, we assess what effect some future directions could have on this study. 
We examine the (preliminary) matching of our small-$x$ gluon hPDF $\Delta g$ onto the larger-$x$ $\Delta g$ coming from a JAM-DGLAP analysis, showing how including $pp$ data now makes the two more compatible. 
We also investigate how EIC data will affect the uncertainties of $g_1^p$.

\subsection{Methodology} \label{ssec:methodology}

\subsubsection{Observables}

Our fit (herein referred to as ``JAMsmallx'') includes world data at small $x$ on double-longitudinal spin asymmetries in DIS, SIDIS, and $pp$ collisions.
As in our previous global analysis~\cite{Adamiak:2023yhz}, the  DIS asymmetries $A_{\parallel}$, $A_1$ and the SIDIS asymmetry $A_1^h$ at small $x$ are approximated as
\begin{equation}
     A_{\parallel} \approx A_1 \approx \frac{g_1(x,Q^2)}{F_1(x,Q^2)}\,,\quad\quad A_1^h \approx \frac{g_1^h(x,z_h,Q^2)}{F_1^h(x,z_h,Q^2)}\,,
\end{equation}
where $g_1^{(h)}$ and $F_1^{(h)}$ are the polarized and unpolarized (SI)DIS structure functions, respectively. The double-longitudinal spin asymmetry $A^{\rm jet}_{LL}$ for single-inclusive jet production in proton-proton collisions is given by
\begin{equation}\label{ALL_ratio}
    A_{LL}^{\mathrm{jet}} \equiv \frac{(\dd\Delta\sigma/\dd^2p_T\, \dd y)^{pp\to \mathrm{jet} X}}{(\dd\sigma/\dd^2p_T\, \dd y)^{pp\to \mathrm{jet} X}}\,.
\end{equation}

Using the expressions for $\Delta q^\pm$, $g_1$, and $g_1^h$ in Eqs.~\eqref{hPDFs_log} and \eqref{g1_log} and $(\dd \Delta\sigma/\dd^2p_T\dd y)^{pp\to \mathrm{jet} X}$ in Eq.~(\ref{pJet_GA}), we can analyze the experimental data within our small-$x$ helicity evolution framework. We fit our basis hPDFs to the measurements of DIS, SIDIS, and $A_{LL}^{\mathrm{jet}}$ asymmetries at $x < 0.1$ to determine appropriate initial conditions of the polarized dipole amplitudes \eqref{Dipole_ICs}. From these extracted initial conditions we can employ the small-$x$ helicity evolution equations \eqref{eq_LargeNc+q} to compute hPDFs at smaller values of $x$, and ultimately evolve down and predict hPDFs at arbitrarily small $x$.

The unpolarized denominators for each asymmetry, or specifically the structure functions $F_1$ and $F_1^h$ for DIS and SIDIS and $(\dd\sigma / \dd^2p_T\dd y)^{pp\to \mathrm{jet} X}$ for $pp$, are computed at NLO using collinear factorization and DGLAP evolution as described in their respective JAM analyses \cite{Cocuzza:2022jye} (DIS, SIDIS) and \cite{Zhou:2022wzm} ($pp$). While the unpolarized cross sections have their own small-$x$ evolution \cite{Mueller:1994rr, Mueller:1994jq, Mueller:1995gb, Balitsky:1995ub, Balitsky:1998ya, Kovchegov:1999yj, Kovchegov:1999ua, JalilianMarian:1997dw, JalilianMarian:1997gr, Weigert:2000gi, Iancu:2001ad, Iancu:2000hn, Ferreiro:2001qy}, those effects are sub-leading compared to the effects of helicity evolution at small $x$, because the former is a single-logarithmic evolution resumming powers of $\alpha_s \ln(1/x)$ and the latter is double-logarithmic, resumming powers of $\alpha_s \ln^2(1/x)$. We leave a simultaneous treatment of small-$x$ evolution in both the polarized numerators and unpolarized denominators of the spin asymmetries for future work. For this study the calculations from Refs.~\cite{Cocuzza:2022jye, Zhou:2022wzm} serve us sufficiently well as a faithful proxy of the actual values of the unpolarized denominators. Additionally, we stress that all partonic channels are kept in the unpolarized $pp$ cross section (and not just the pure-glue channel, as in the polarized cross section). Our intent is specifically to identify the systematic impact that $pp$ data have on our small-$x$ extractions of the hPDFs, and this would not be aided by artificially removing contributing channels from the unpolarized denominator. We elect instead to treat the DIS, SIDIS, and $pp$ unpolarized cross sections consistently, and use their unaltered JAM-DGLAP calculations~\cite{Cocuzza:2022jye, Zhou:2022wzm}.

The inclusion of polarized $pp$ jet production data is an important improvement to the small-$x$ helicity global analysis. In Ref.~\cite{Adamiak:2023yhz}, we found that the gluon-dominated polarized dipole amplitudes $\widetilde{G}$ and $G_2$ had the largest influence as $x$ decreased, yet at the same time there were no observables that coupled directly to either of them.\footnote{To be more precise, the basis hPDFs constructed \textit{from} $\widetilde{G}$ and $G_2$ initial conditions were dominant at small $x$. This was first discussed in Ref.~\cite{Adamiak:2023yhz} in the context of the large-$N_c\&N_f$ evolution equations. Nevertheless, this behavior persists in the large-$N_c^{+q}$ evolution equations.} 
The DIS and SIDIS observables only couple directly to $\Delta q^{\pm}$, and from Eq.~\eqref{hPDFs_log} we see that the dependence on $G_2$ only enters in the linear combination $Q_q + 2G_2$ (and not separately for $G_2$). The dipole amplitude $\widetilde{G}$ does not contribute to any observable/hPDF at all except indirectly through small-$x$ evolution. With the addition of $pp$ data we now have an observable with direct sensitivity to $\widetilde{G}$ and $G_2$ (see Eq.~\eqref{pJet_GA}), which should provide a better constraint on the small-$x$ behavior in the kinematic region where these dipole amplitudes dominate.

Using the JAM Bayesian Monte Carlo framework \cite{Sato:2016tuz, Sato:2019yez, Moffat:2021dji}, we  sample the 24 initial condition parameters (the $a$'s, $b$'s, and $c$'s from Eq.~\eqref{Dipole_ICs}) and compute an evolved set of polarized dipole amplitudes constructed from the specific combination of basis functions. Note that our definition of basis functions~\cite{Adamiak:2023yhz} implies that we have already solved the evolution equations \eqref{eq_LargeNc+q}, and are simply combining the \textit{evolved} set of basis polarized dipole amplitudes, each being multiplied by their respective parameters, to obtain the dipole amplitudes corresponding to a given set of initial conditions. These constructed polarized dipole amplitudes are then used to compute $g_1$, $g_1^h$, and $(\dd \Delta\sigma/\dd^2p_T\dd y)^{pp\to \mathrm{jet} X}$ via Eqs.~\eqref{pJet_GA} and \eqref{g1_log}, which determine the numerators of the spin asymmetries. From a random sampling of initial parameters, a gradient descent $\chi^2$-minimization algorithm finds the optimized set of parameters from which we can reconstruct the polarized dipole amplitudes, and subsequently the hPDFs, and predict their values at small $x$. In order to propagate the uncertainty of the experimental data, we employ data reshuffling. This is the process of generating ``pseudodata" from sampling within the uncertainty of the actual data. When the $\chi^2$ is minimized with respect to these pseudodata, that generates one set of theory parameters which we refer to as a ``replica.'' This process is repeated 500 times to provide a statistically stable set of results.

Lastly, we note an update regarding the previous large-$N_c\&N_f$ small-$x$ helicity DIS and SIDIS analysis~\cite{Adamiak:2023yhz}. 
For those observables we require input from the unpolarized structure functions $F_1$ and $F_1^h$ and hadronic fragmentation functions $D_q^h$, which are not small-$x$ specific objects. These fragmentation functions were previously taken from the JAM analysis in Ref.~\cite{Cocuzza:2022jye}. Since then it has been pointed out in Ref.~\cite{Anderson:2024evk} that for one source of data used in extracting these functions, the systematic uncertainty was improperly taken as a normalization uncertainty. This was corrected in Ref.~\cite{Anderson:2024evk}, and we use those updated fragmentation functions, in addition to imposing new kinematic cutoffs that will be mentioned below.

\subsubsection{Cuts on data}\label{data_cuts}

We use the same data for DIS and SIDIS as Ref.~\cite{Adamiak:2023yhz}.
The measurements go down to $x = 5\times 10^{-3}$, and we place an upper $x$ cutoff of 0.1, a choice which was justified in Refs.~\cite{Adamiak:2021ppq,Adamiak:2023yhz}. For $5\times 10^{-3}< x < 0.1$, the data have a maximum $Q^2$ of $10.4~\mathrm{GeV}^2$, and we place a lower $Q^2$ cut of $m_c^2=1.69~\mathrm{GeV}^2$ (the same cutoff for the set of fragmentation functions we use~\cite{Anderson:2024evk}). As described above, and to remain consistent with previous small-$x$ helicity global analyses \cite{Adamiak:2021ppq, Adamiak:2023yhz}, we choose $\Lambda = 1~\mathrm{GeV}$ without loss of generality since any changes in $\Lambda$ can be absorbed by redefining the $c$ parameters in Eqs.~\eqref{Dipole_ICs}. For the polarized SIDIS data, we use the cuts $W_{\mathrm{SIDIS}}^2 > 10~\mathrm{GeV}^2$ and $z_h < 0.8$. Going forward, all results we present are understood to have made these changes to the fragmentation functions and SIDIS data cuts, including retroactive updates to the large-$N_c\&N_f$ analysis of DIS and SIDIS from Ref.~\cite{Adamiak:2023yhz}.

For inclusive jet production data from polarized $pp$ collisions, we have to be more careful in determining what constitutes ``small $x$". 
The momentum fractions of the (massless) partons from the projectile and target protons are $x_{T,P}= \tfrac{p_T}{\sqrt{s}}e^{\pm y}$, where $p_T$ and $y$ are the transverse momentum and rapidity of the produced jet, and $\sqrt{s}$ is the CM energy.
To be consistent with the assumption made in Ref.~\cite{kovchegov:2024aus} in deriving \eq{KLprod} that the gluon is produced at central rapidity values, we conservatively allow only data where both $x_T, x_P$ are small $(x_T, x_P < 0.1)$ across the entire $y$ and $p_T$ range of a given bin (as opposed to using their central values). 
In a manner similar to the approach in Ref.~\cite{Adamiak:2021ppq}, we tested the applicability of our small-$x$ helicity evolution framework to various small-$x$ cutoffs for polarized $pp$ data and found that both the ability of a given fit to converge and the resultant $\chi^2/N_{\mathrm{pts}}$ of a converged fit degrades for cutoffs of $x_0 > 0.1$.
The $\chi^2/N_{\mathrm{pts}}$ values of fits to $pp$ data with different small-$x$ cutoffs were found to systematically increase as $x_0$ grew larger. 
For example, small-$x$ cutoffs of $x_0 = 0.125$ and $x_0 = 0.15$ had $\chi^2/N_{\mathrm{pts}}$ of $0.71$ and $1.08$, respectively, which are roughly two times and three times larger than the $\chi^2/N_{\mathrm{pts}}$ for the $x_0 = 0.1$ baseline fit discussed below.

Furthermore, we have to ensure the $A_{LL}^{\rm jet}$ data we use is consistent with the assumptions of our polarized $pp$ jet production cross section, and that it remains compatible with the unpolarized  cross section calculated in Ref.~\cite{Zhou:2022wzm}. 
The accuracy of the NJA discussed above was assessed in Ref.~\cite{Jager:2004jh}, which compared the NJA with the full jet cross section using a Monte Carlo simulation \cite{deFlorian:1998qp}. 
They found that the NJA was accurate to within $\sim 5- 10\%$ for the polarized cross section (and significantly better for the unpolarized cross section) for $R \leq 0.7$. 
This result informs our choice of selection cuts on jet production data based on algorithm and on $R$.
Therefore, we disregard one available dataset with $R = 1$ and proceed with the remaining measurements of $A_{LL}^{\mathrm{jet}}$, which all have jet radii $R \leq 0.7$. 
In addition, the JAM analysis~\cite{Zhou:2022wzm} used for the unpolarized cross section places a cut of $p_T\geq 8\,{\rm GeV}$ on the data, so we also impose this restriction on~$A^{\rm jet}_{LL}$. 
In the end we are left with 122 polarized DIS data points, 104 polarized SIDIS data points, and 14 $pp$ data points that survive the aforementioned cuts, making a total of $N_{\mathrm{pts}} = 240$ data points.

\subsection{Data versus theory comparisons}\label{ssec:data_vs_thy}

The world polarized DIS, SIDIS, and $pp$ single-inclusive jet production data that we have access to for this analysis comes from a large variety of colliders and collaborations. 
The DIS measurements are from COMPASS \cite{Alekseev:2010hc, Adolph:2015saz, Adolph:2016myg}, EMC \cite{Ashman:1989ig}, HERMES \cite{Ackerstaff:1997ws, Airapetian:2007mh}, SLAC \cite{Anthony:1996mw, Abe:1997cx, Abe:1998wq, Anthony:1999rm, Anthony:2000fn}, and SMC \cite{Adeva:1998vv, Adeva:1999pa}, while the SIDIS measurements are from COMPASS \cite{COMPASS:2009kiy, COMPASS:2010hwr}, HERMES \cite{HERMES:1999uyx, HERMES:2004zsh}, and SMC \cite{SpinMuon:1998eqa}. 
The $pp$ data are from four different measurements from STAR, Refs.~\cite{STAR:2014wox, STAR:2019yqm, STAR:2021mfd, STAR:2021mqa}.
Tables \ref{t:Chi2_DIS}, \ref{t:Chi2_SIDIS}, and \ref{t:Chi2_pJet} provide a breakdown of individual datasets used in this analysis, accompanied by their respective $\chi^2/N_{\mathrm{pts}}$. We continue to use the same $\chi^2$ formula employed in other JAM-based analyses,
\begin{equation}
    \chi^2 = \sum_{e,i}\bigg(\frac{d_{e,i} - \mathrm{E}_{e,i}^{\mathrm{thy}}}{\alpha_{e,i}}\bigg)^{\! 2},
\end{equation}
where $d_{e,i}$ refers to a specific experimental value from dataset $e$ and data point $i$, and $\mathrm{E}_{e,i}^{\mathrm{thy}}$ is the mean theory value of all replicas, shifted and normalized by the correlated uncertainties of the relevant dataset/data point, and $\alpha_i$ is the uncorrelated uncertainty for data point $i$. 

\begin{table}[b!]
  \caption{Summary of polarized DIS $A_1$ (left) and $A_{\parallel}$ (right) data included in the analysis, along with the $\chi^2/N_{\rm pts}$ values for each dataset.\\}
  \label{t:Chi2_DIS}
    \begin{tabular}{l|c|c|c} 
    \hline
    {\bf Dataset ($\boldsymbol{A_1}$)} & 
        ~{\bf Target}~ &
        ~$\boldsymbol{N_\mathrm{pts}}$~ & 
        ~$\boldsymbol{\chi^2 / N_\mathrm{pts}}$~  
        \\ \hline     
    COMPASS   \cite{Alekseev:2010hc} &
        $p$ &
        $5$ &
        $0.77$  \\ \hline
    COMPASS   \cite{Adolph:2015saz} &
        $p$ &
        $17$ &
        $0.93$   \\ \hline
    COMPASS  \cite{Adolph:2016myg} &
        $d$ &
        $5$ &
        $0.34$  \\ \hline
    EMC  \cite{Ashman:1989ig} &
        $p$ &
        $5$ &
        $0.23$   \\ \hline
    HERMES   \cite{Ackerstaff:1997ws} &
        $n$ &
        $2$ &
        $1.11$ \\ \hline
    SLAC (E142) \cite{Anthony:1996mw} &
        ${}^3 \mathrm{He}$ &
        $1$       & 
        $1.47$    \\ \hline
    SMC  \cite{Adeva:1998vv, Adeva:1999pa} &
        $p$ &
        $6$ &
        $1.26$  \\ 
        &
        $p$ &
        $6$ &
        $0.43$  \\
        &
        $d$ &
        $6$ &
        $0.65$  \\
        &
        $d$ &
        $6$ &
        $2.13$  	\\ \hline\hline
    {\bf Total} &    & 59 & 0.90 \\ \hline
   \end{tabular}
    \qquad\qquad\qquad
    \begin{tabular}{l|c|c|c} 
    \hline
    {\bf Dataset ($\boldsymbol{A_{\parallel}}$)}& 
        ~{\bf Target}~ &
        ~$\boldsymbol{N_\mathrm{pts}}$~ & 
        ~$\boldsymbol{\chi^2 / N_\mathrm{pts}}$~  
        \\ \hline
    HERMES    \cite{Airapetian:2007mh} &
        $p$ &
        $4$ &
        $1.47$  \\ 
        &
         $d$ &
         $4$ &
         $1.00$  \\ \hline
    SLAC (E143) \cite{Abe:1998wq} &
        $p$ &
        $9$ &
        $0.55$    \\
        &
        $d$ &
        $9$ &
        $1.01$  	\\ \hline
    SLAC (E154) \cite{Abe:1997cx} &
        ${}^3 \mathrm{He}$ &
        $5$       &
        $0.69$     \\ \hline
    SLAC (E155) \cite{Anthony:1999rm} &
        $p$ &
        $16$ &
        $1.07$ 	\\ 
        &
        $d$ &
        $16$ &
        $1.57$      \\ \hline\hline
    {\bf Total} &   & 63 & 1.10 \\ \hline
    \end{tabular}
  \label{t:Chi2_DIS_Apa}
\end{table}

\begin{table}[t!]
    \begin{center}
    \caption{Summary of the polarized SIDIS data on $A_1^h$ used in the present analysis, along with the $\chi^2/N_{\rm pts}$ for each dataset.\\}
    \label{t:Chi2_SIDIS}
    \vspace{0.3cm}
    \begin{tabular}{l|c|c|c|c} 
    \hline
    {\bf Dataset ($\boldsymbol{A_1^h}$})& 
        ~{\bf Target}~ &
        ~{\bf Tagged Hadron}~ &
        ~$\boldsymbol{N_\mathrm{pts}}$~ & 
        ~$\boldsymbol{\chi^2 / N_\mathrm{pts}}$~  
        \\ \hline
    COMPASS \cite{COMPASS:2009kiy}  
        &
        $d$ &
        $\pi^+$ &
        $5$   &
        $0.63$ 	\\ 
        &
        $d$  &
        $\pi^-$ &
        $5$   &
        $0.83$ \\
        &
        $d$   &
        $h^+$ &
        $5$ &
        $1.01$ \\ 
        &
        $d$ &
        $h^-$  &
        $5$  &
        $1.02$ 		\\ 
        &
        $d$     &
        $K^+$ &
        $5$  &
        $1.60$ 		\\
        &
        $d$ &
        $K^-$ &
        $5$ &
        $0.71$ 	\\ \hline
    COMPASS \cite{COMPASS:2010hwr} &
        $p$ &
        $\pi^+$ &
        $5$ &
        $1.94$ 		\\ 
        &
        $p$   &
        $\pi^-$ &
        $5$ &
        $1.18$ \\ 
        &
        $p$  &
        $K^+$ &
        $5$   &
        $0.46$ 	\\
        &
        $p$ &
        $K^-$ &
        $5$   &
        $0.23$ 		\\ \hline	
    HERMES \cite{HERMES:1999uyx} &
        ${}^3 \mathrm{He}$  &
        $h^+$  &
        $2$ &
        $0.55$ 	\\ 
        &
        ${}^3 \mathrm{He}$ &
        $h^-$  &
        $2$   &
        $0.29$ 	\\ \hline
    HERMES  \cite{HERMES:2004zsh} &
        $p$ &
        $\pi^+$ &
        $2$ &
        $2.75$ \\ 
        &
        $p$ &
        $\pi^-$ &
        $2$  &
        $0.00$    \\ 
        &
        $p$ &
        $h^+$ &
        $2$ &
        $1.25$  \\
        &
        $p$ &
        $h^-$ &
        $2$  &
        $0.19$  \\ 
        &
        $d$  &
        $\pi^+$ &
        $2$ &
        $0.58$ 	 \\
        &
        $d$ &
        $\pi^-$ &
        $2$ &
        $1.23$ 	 \\
        &
        $d$ &
        $h^+$  &
        $2$  &
        $3.03$   \\ 
        &
        $d$  &
        $h^-$ &
        $2$  &
        $1.24$ 	 \\ 
        &
        $d$ &
        $K^+$ &
        $2$  &
        $0.82$ 	 \\ 
        &
        $d$  &
        $K^-$ &
        $2$  &
        $0.25$ 	\\ 
        &
        $d$  &
        $K^+ + K^-$ &
        $2$  &
        $0.36$ 	\\ \hline
    SMC  \cite{SpinMuon:1998eqa} &
        $p$ &
        $h^+$ &
        $7$ &
        $1.22$  \\ 
        &
        $p$ &
        $h^-$ &
        $7$ &
        $1.41$  		\\ 
        &
        $d$ &
        $h^+$ &
        $7$ &
        $0.84$  \\ 
        &
        $d$ &
        $h^-$ &
        $7$ &
        $1.52$   \\ \hline\hline
    {\bf Total} &    &    & 104 & 1.04 \\ \hline
    \end{tabular}
    \end{center}
\end{table}

\begin{table}[t!]
  \caption{Summary of polarized $pp$ data for $A_{LL}^{\mathrm{jet}}$ included in our analysis, along with the  $\chi^2/N_{\rm pts}$ values.\\}
  \label{t:Chi2_pJet}
    \vspace{0.3cm}
    \begin{tabular}{l|c|c} 
    \hline
    {\bf Dataset ($\boldsymbol{A_{\mathrm{LL}}^{\mathrm{jet}}}$)}& 
        ~$\boldsymbol{N_\mathrm{pts}}$~ & 
        ~$\boldsymbol{\chi^2 / N_\mathrm{pts}}$~  
        \\ \hline
    STAR    \cite{STAR:2014wox} &
        $2$ &
        $0.60$  \\ \hline
    STAR \cite{STAR:2019yqm} &
        $5$ &
        $0.30$    \\ \hline
    STAR \cite{STAR:2021mfd} &
        $2$       &
        $0.55$     \\ \hline
    STAR \cite{STAR:2021mqa} &
        $5$ &
        $0.24$ 	\\ \hline\hline
    {\bf Total} & 14 & 0.36 \\ \hline
    \end{tabular}
\end{table}

\begin{figure}[h!] 
\begin{centering}
\includegraphics[width=400 pt]{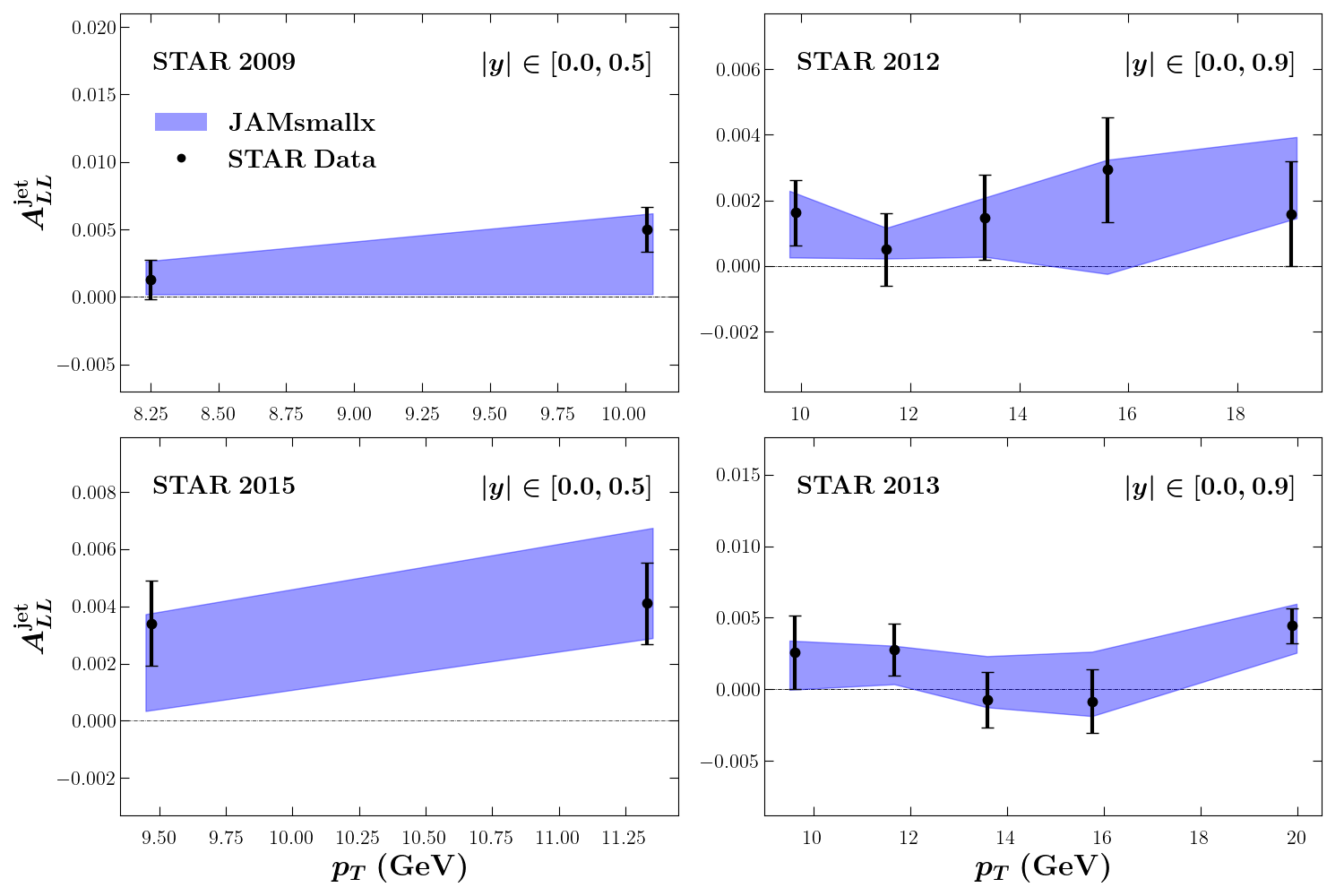}
\caption{Jet asymmetry $A_{LL}^{\mathrm{jet}}$ data from STAR~\cite{STAR:2014wox, STAR:2019yqm, STAR:2021mfd, STAR:2021mqa} (black circles) as a function of $p_T$ in the kinematic range $p_T \geq 8$~GeV, $-1 < y < +1$, and $x_{T,P} < 0.1$, compared with the fitted asymmetries (blue bands, 68\% confidence intervals).
    \label{pjet_data_vs_thy_2smx}
}
\end{centering}
\end{figure}

%
In Fig.~\ref{pjet_data_vs_thy_2smx} we show our theory curves for $A_{LL}^{\mathrm{jet}}$, which give a very good description of the data, although there are a limited number of points. 
The comparison of data and theory for DIS and SIDIS are similar to what was found in Ref.~\cite{Adamiak:2021ppq}, so we do not show those plots here. 
We find that the overall $\chi^2/N_{\mathrm{pts}}$ is 0.98. Thus, we have been able to incorporate $pp$ data into our global analysis without degrading our ability to fit the DIS and SIDIS data. 
Indeed, we find that the $\chi^2/N_{\mathrm{pts}}$ for the DIS and SIDIS data together is 1.02, which matches the DIS+SIDIS only fit of Ref.~\cite{Adamiak:2023yhz}, while for the $pp$ data it is 0.36.

The results are based on a set of 398 replicas that passed a ``filter'' of $\chi^2/N_{\mathrm{pts}} \leq 10$ per reaction (DIS, SIDIS, and $pp$), with the fit being iterated a second time to ensure convergence. 
That is, we took the final replicas of the first fit and re-fit them to the same data, further optimizing the overall $\chi^2/N_{\mathrm{pts}}$ of each replica.
This second iteration of fitting saw a $1.04\%$ improvement to the first fit's $\chi^2/N_{\mathrm{pts}}$ value, while a subsequent third iteration only improved the $\chi^2/N_{\mathrm{pts}}$ by $0.18\%$, indicating the second iteration had already sufficiently converged.
The results we present are those that came from the second-iteration fit.

\subsection{Effect of $pp$ data on $g_1$ and hPDFs} \label{ssec:effectofpp}

\begin{figure}[t]
\begin{centering}
\includegraphics[width=500 pt]{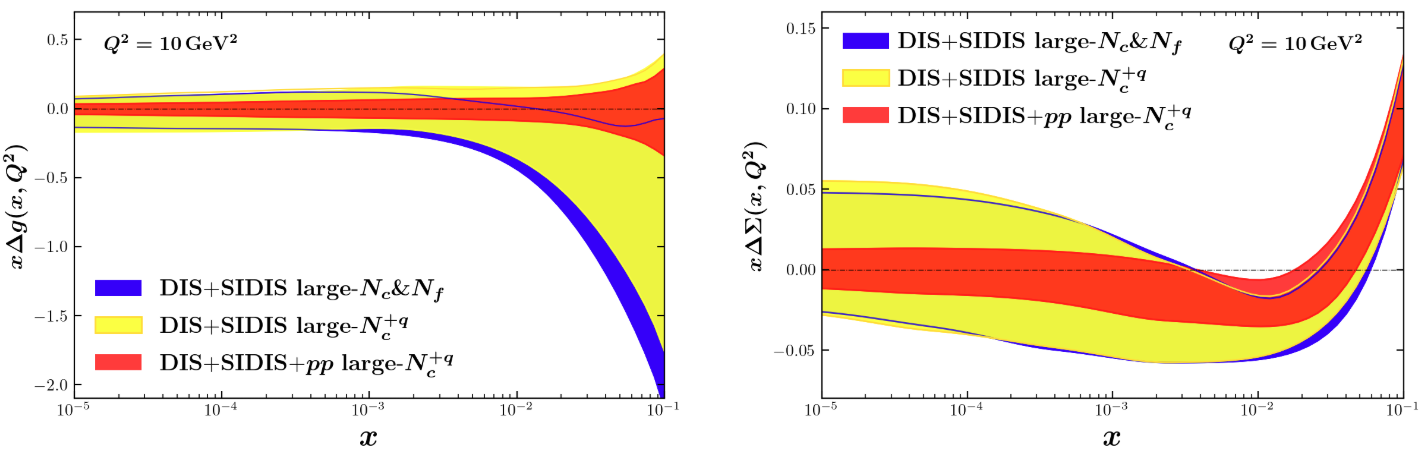}
\caption{The gluon hPDF $x\Delta g(x,Q^2)$ (left) and flavor singlet quark hPDF combination $x\Delta\Sigma(x,Q^2)$ (right) as a function of $x$ at $Q^2 = 10$~GeV$^2$. The bands represent 68\% confidence intervals from three analyses:~DIS+SIDIS data with the large-$N_c\&N_f$ evolution equations (blue), DIS+SIDIS data with the large-$N_c^{+q}$ evolution equations (yellow), and DIS+SIDIS+$pp$ data with the large-$N_c^{+q}$ evolution equations (red). 
    \label{DeltaG_DeltaSigma}
}
\end{centering}
\end{figure}

\begin{figure}[h!]
\begin{centering}
\includegraphics[width=500 pt]{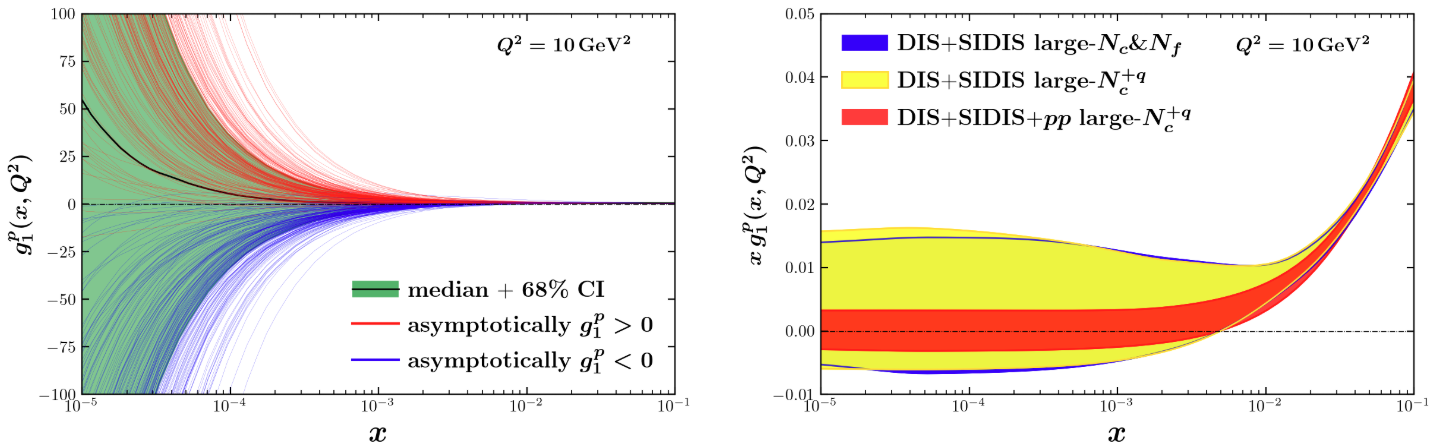}
\caption{(Left) $g_1^p(x, Q^2)$ structure function versus $x$ at $Q^2=10\,{\rm GeV^2}$ for the DIS+SIDIS+$pp$ analysis with the large-$N_c^{+q}$ evolution equations, showing replicas that grow asymptotically positive (red) and negative (blue) as $x\to 0$ (measured at $x=10^{-7}$). The black curve represents the median value of all replicas while the green band shows the 68\% confidence interval of those same replicas.  
(Right) $x \,g_1^p(x,Q^2)$ versus $x$ at $Q^2 = 10~\mathrm{GeV}^2$ for the DIS+SIDIS data with the large-$N_c\&N_f$ evolution equations (blue), DIS+SIDIS data with the large-$N_c^{+q}$ evolution equations (yellow), and DIS+SIDIS+$pp$ data with the large-$N_c^{+q}$ evolution equations (red). These bands also reflect the 68\% confidence intervals of all replicas for their respective analyses.
    \label{g1_band_comparison}
}
\end{centering}
\end{figure}

The results of our current and previous analyses for the hPDFs $\Delta g(x,Q^2)$ and $\Delta\Sigma(x,Q^2)$ and the $g_1^p(x,Q^2)$ structure function are depicted in Figs.~\ref{DeltaG_DeltaSigma} and \ref{g1_band_comparison} as a function of $x$.
The bands on the helicity distributions (weighted by $x$) in Fig.~\ref{DeltaG_DeltaSigma} represent 68\% confidence level uncertainties. 
The plots compare three different analyses that differ in their observables and/or evolution equations:~DIS and SIDIS data using large-$N_c\&N_f$ evolution; DIS and SIDIS data using large-$N_c^{+q}$ evolution; and DIS, SIDIS, and $pp$ data using large-$N_c^{+q}$ evolution. 
The new distribution of replicas for $g_1^p$ when fitting to $pp$ data is shown in Fig.~\ref{g1_band_comparison} as a function of $x$, along with a comparison of the three analyses described above for the small-$x$ behavior of $x g_1^p$. 
To identify the influence of the $pp$ data on our results, we must first make clear how the DIS+SIDIS analysis differs as a result of changing between the large-$N_c\&N_f$ and large-$N_c^{+q}$ evolution equations. 
To do so, we repeated the analysis from Ref.~\cite{Adamiak:2023yhz}, fitting to the same DIS and SIDIS data, but this time solving the evolution equations~\eqref{eq_LargeNc+q}. We highlight the important distinctions below.

First, we note that the goodness-of-fit does not degrade. 
From the large-$N_c\&N_f$ global analysis of DIS and SIDIS data to the large-$N_c^{+q}$ one, we see a negligible change in the $\chi^2/N_{\mathrm{pts}}$ from 1.017 to 1.022. 
Considering that both analyses can describe the same data equally well, we can now discuss the differences made to the overall uncertainty of our results and the differences in the quark ($C$-even) and gluon hPDFs and the $g_1^p$ structure function.

The change in the evolution equations does not affect the observation that the hPDF basis functions constructed from the $\widetilde{G}$ and $G_2$ initial conditions make the largest contributions at small $x$. 
The small-$x$ behavior of $\widetilde{G}$ and $G_2$ result in two important phenomenological features:~{\bf (1)}~the small-$x$ asymptotics of $g_1$, $\Delta \Sigma$, and $\Delta g$ are correlated, and {\bf (2)}~the uncertainty at small $x$ is largely controlled by the constraining power of the data on the $\widetilde{G}$ and $G_2$ polarized dipole amplitudes~\cite{Adamiak:2023yhz}. 
Both of these statements remain true upon changing the evolution equations. 
However, their effect on the results are subtly different. The overall split of $g_1^p$ replicas in the large-$N_c^{+q}$ analysis of DIS and SIDIS data has $69\%$ of replicas trending asymptotically positive, comparable to the $70\%$ positive trend seen in Ref.~\cite{Adamiak:2023yhz}.
Although the change in evolution equations does not significantly affect the balance of asymptotically positive and negative solutions, it does change the \textit{shape} of those replicas. 
The resultant changes to the helicity distributions can be seen in Fig.~\ref{DeltaG_DeltaSigma}. 
The gluon hPDF $\Delta g$ has now shifted more positive near the boundary $x \lesssim 0.1$.

In going from the large-$N_c\&N_f$ to the large-$N_c^{+q}$ evolution equations, we see from Fig.~\ref{g1_band_comparison} that without the extra constraint from including $N_f$-enhanced terms the small-$x$ uncertainty in $g_1^p(x,Q^2)$ is slightly larger (blue versus yellow band in the right panel of Fig.~\ref{g1_band_comparison}). 
Using either the large-$N_c\&N_f$ or the large-$N_c^{+q}$ evolution equations, the theory can describe the experimental measurements equally well in the range $5\times10^{-3}< x< 0.1$, but once the data run out, the replicas (left panel of Fig.~\ref{g1_band_comparison}) move towards their asymptotics slightly quicker in the large-$N_c^{+q}$ case. 
This extends to $\Delta\Sigma(x,Q^2)$ and $\Delta g(x,Q^2)$, where we find the small-$x$ uncertainty is slightly larger when fitting with the large-$N_c^{+q}$ evolution equations, although the overall behavior of these functions remains consistent between the blue and yellow bands --- see Fig.~\ref{DeltaG_DeltaSigma}. 
In agreement with the anticorrelation~\cite{Adamiak:2023yhz}, 
\begin{equation}\label{hPDF_correlation}
    \lim_{x\to 0} g_1(x,Q^2) \sim \lim_{x\to 0} \Delta \Sigma(x,Q^2) \sim \lim_{x\to 0} -\Delta g(x,Q^2)\,,
\end{equation}
the slight positive mean value as $x\to 0$ for $\Delta\Sigma$ (yellow band, right panel) is accompanied by a slightly negative mean value as $x\to 0$ for $\Delta g$ (yellow band, left panel).


The effect of using the large-$N_c^{+q}$ evolution equations can then be summarized as slightly increasing the uncertainty bands and shifting the gluon hPDF $\Delta g(x,Q^2)$ towards slightly more positive values near $x = 0.1$, while retaining its overall preference for asymptotically negative solutions. 
The correlations between $g_1$ and the hPDFs is unaffected, as is the small-$x$ dominance of the basis hPDFs constructed from the $\widetilde{G}$ and $G_2$ polarized dipole amplitude initial conditions. 
Any further changes to our small-$x$ extractions are thus caused strictly by including polarized $pp$ data.

The polarized proton structure function $g_1^p(x,Q^2)$, depicted in Fig.~\ref{g1_band_comparison} (left panel), at first glance it looks remarkably similar to its counterpart in Fig.~5 of Ref.~\cite{Adamiak:2023yhz}. 
The asymptotic trend of the replicas towards $\pm \infty$ is a feature of the DLA evolution equations (see Eq.~\eqref{asymptotics}). 
This bimodality of solutions was attributed in Ref.~\cite{Adamiak:2023yhz} to the insensitivity of data to the specific signs of the $\widetilde{G}$ and $G_2$ polarized dipoles amplitudes.
After the addition of $pp$ data, we found that the insensitivity is less discriminant towards asymptotically negative $g_1^p$ solutions, where we now have a more even split, with only 54\% of replicas growing asymptotically positive (compared to 70\% in Ref.~\cite{Adamiak:2023yhz}). 
This persistent bimodality results in a rather large uncertainty band centered near zero. 
However, the crucial difference with this analysis compared to Ref.~\cite{Adamiak:2023yhz} is that $pp$ data provide overall better constraint on the parameters for the $\widetilde{G}$ and $G_2$ initial conditions. 
This causes significantly smaller uncertainty bands for the hPDF $\Delta g$, $\Delta\Sigma$ and the structure function $g_1^p$ than the extraction without $pp$ data, as shown in Figs.~\ref{DeltaG_DeltaSigma} and \ref{g1_band_comparison}.
Quantitatively we determined that at $x = 10^{-7}$ and $Q^2=10$~GeV$^2$, the $g_1^p(x,Q^2)$ uncertainty band of the large-$N_c^{+q}$ analysis of only DIS and SIDIS data is reduced by 73\% when polarized $pp$ data is included. 
\footnote{We have chosen to compute small-$x$ extractions only down to $x=10^{-7}$ because this is the approximate prediction for $x_{\mathrm{sat}}$, the value of $x$ at which saturation effects set in at a given $Q^2$, as determined by using the Golec-Biernat Wusthoff parametrization for the proton saturation scale \cite{Golec-Biernat:1998zce} at $Q^2=10~\mathrm{GeV}^2$: we required that $Q_{s,p}^2 (x_{\rm sat}) = Q^2=10~\mathrm{GeV}^2$ and solved for $x_{\rm sat}$, obtaining $x_{\rm sat} \approx 10^{-7}$.} 
Measurements of the uncertainty at extremely small values of $x$ are informative, but the EIC will only provide data as low as $x = 10^{-4}$.
We found that at this value of $x$ the uncertainty band for $g_1^p(x,Q^2)$ is reduced by a very similar 69\% when comparing the DIS+SIDIS+$pp$ analysis to the DIS+SIDIS analysis.
These reductions in our small-$x$ uncertainty are a direct result of the constraining power of $pp$ data on the $\widetilde{G}$ and $G_2$ initial conditions, as evidenced by the standard deviation of the three $\widetilde{G}$ parameters being reduced on average by 52\% and the three $G_2$ parameters by 23\% compared to the DIS+SIDIS large-$N_c^{+q}$ analysis.

Clearly the incorporation of polarized $pp$ data has a strong effect on the gluon hPDF $\Delta g$, and by extent the overall helicity contribution from small-$x$ partons. 
Going back to Fig.~\ref{DeltaG_DeltaSigma}, we find that the sensitivity of $pp$ data to the gluon-dominated dipole amplitudes provides a $73\%$ reduction in the uncertainty of $\Delta g$ at $x = 10^{-7}$ and $Q^2 = 10~\mathrm{GeV}^2$ (with a similar reduction found in $\Delta\Sigma$). 
The large reduction in uncertainty remains across all values of $x<0.1$ when compared against the DIS+SIDIS only analysis in the large-$N_c^{+q}$ limit, and the resulting $\Delta g(x,Q^2)$ strictly satisfies the positivity constraint. 
The other most noticeable change in $\Delta g$ is the significant reduction in negative $\Delta g$ replicas for $0.01\lesssim x < 0.1$, suggesting that polarized $pp$ data discourages large $|\Delta g|$.
We can further explore the differences with the DIS+SIDIS only fit via truncated integrals over $x$ of $\Delta g(x, Q^2)$ and $\Delta\Sigma(x,Q^2)$.

\begin{figure}[t!]
\begin{centering}
\includegraphics[width=300pt]{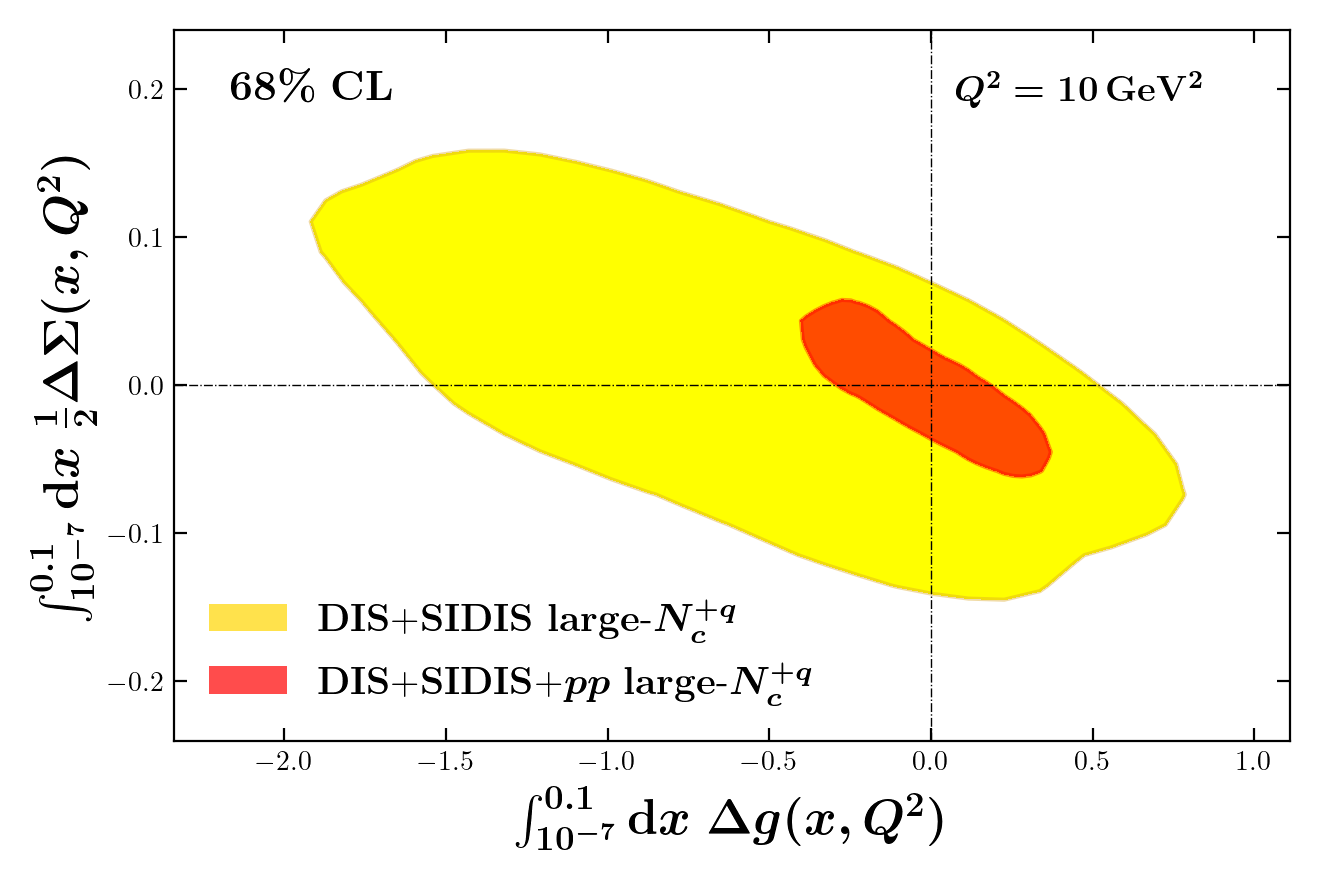}
\caption{Truncated integrals of $\frac{1}{2}\Delta\Sigma(x,Q^2)$ and $\Delta g(x, Q^2)$ over the range $x\in[10^{-7}, 0.1]$ for the DIS+SIDIS large-$N_c^{+q}$ (yellow) and DIS+SIDIS+$pp$ large-$N_c^{+q}$ (red) analyses, at $Q^2=10~\mathrm{GeV}^2$. The contours contain 68\% of the replicas.
    \label{xmin_integral_comparison}
}
\end{centering}
\end{figure}

The total small-$x$ helicity contribution from the gluons, $\Delta G(Q^2)$, and all quarks, $\tfrac{1}{2}\Delta\Sigma(Q^2)$, integrated over the range $10^{-7} < x < 0.1$ is shown in Fig.~\ref{xmin_integral_comparison} at $Q^2=10$~GeV$^2$.
The impact on these quantities is from $pp$ data removing a significant number replicas with negative helicity contribution (especially for $\Delta g(x,Q^2)$ near $x \lesssim 0.1$), and reducing the small-$x$ uncertainty for both (see Fig.~\ref{DeltaG_DeltaSigma}). 
This causes the small-$x$ total parton helicity contribution from the region $10^{-7} < x < 0.1$ to now be compatible with zero. 
This outcome is consistent with the correlation~\eqref{hPDF_correlation} and the observation of a much more even split of asymptotically positive and negative $g_1^p(x, Q^2)$ solutions. 
Quantitatively, the small-$x$ gluon and total parton helicity contributions (at $Q^2 = 10~\mathrm{GeV}^2$) for the DIS+SIDIS only analysis (Eq.~\eqref{e:spinDIS}) and the DIS+SIDIS+$pp$ analysis (Eq.~\eqref{e:spinpp}) are
\begin{equation} \label{e:spinDIS}
    \int\limits_{10^{-7}}^{0.1}\,\dd x \, \Delta g (x, Q^2)\Big|_{\mathrm{DIS+SIDIS}} \approx -0.52 \pm 0.96 \,, \qquad \int\limits_{10^{-7}}^{0.1}\,\dd x \, \Bigl(\frac{1}{2}\Delta\Sigma + \Delta g \Bigr)(x, Q^2)\Big|_{\mathrm{DIS+SIDIS}} \approx -0.52 \pm 0.90 \,,
\end{equation}
\begin{equation} \label{e:spinpp}
    \int\limits_{10^{-7}}^{0.1}\,\dd x \,  \Delta g (x, Q^2)\Big|_{+\,pp} \approx -0.04 \pm 0.26\,, \qquad \int\limits_{10^{-7}}^{0.1}\,\dd x \, \Bigl(\frac{1}{2}\Delta\Sigma + \Delta g \Bigr)(x, Q^2)\Big|_{+\,pp} \approx -0.04 \pm 0.23\,,
\end{equation}
where one clearly notices the shift towards zero and the reduced uncertainty bands, which nevertheless are still sizable.

\begin{figure}[t!]
\begin{centering}
\includegraphics[width=525pt]{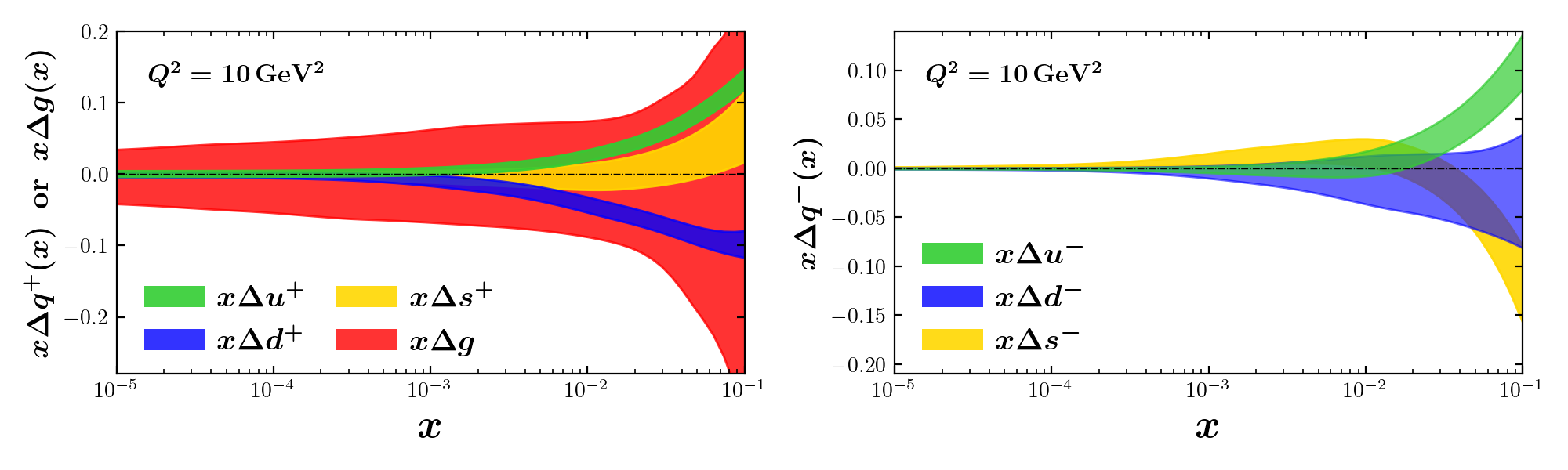}
\caption{(Left) $C$-even hPDFs $x\Delta u^+$, $x\Delta d^+$, and $x\Delta s^+$ and the gluon hPDF $x\Delta g$. (Right) $C$-odd hPDFs $x\Delta u^-$, $x\Delta d^-$, and $x\Delta s^-$. All distributions are at a fixed $Q^2=10$~GeV$^2$. \label{hPDFs_2smx}
}
\end{centering}
\end{figure}

For completeness, the ($x$-scaled) $C$-even $\Delta q^+$ and $C$-odd $\Delta q^-$ quark hPDFs, along with $\Delta g$, are plotted as functions of $x$ in Fig.~\ref{hPDFs_2smx}. 
Note again that there are several differences between our current analysis and the previous small-$x$ hPDF extraction in Ref.~\cite{Adamiak:2023yhz}. These include a change to the evolution equations, a change in the scheme (``polarized DIS" vs.~$\overline{\mathrm{MS}}$), and the inclusion of $pp$ data.
We have discussed the effect of these on $\Delta g$, but there are indeed noticeable modifications to the quark hPDFs as well. 
We investigated the effects due to each of the three changes and found that (at $Q^2 =10~\mathrm{GeV}^2$) $\Delta u^{\pm}$ and $\Delta d^{\pm}$ have a slight reduction in uncertainty and a small decrease in magnitude as $x\to 0.1$, especially for the former, which are attributed to both the change in scheme and the inclusion of $pp$ data.
The difference in $\Delta s^{\pm}$ is more significant. We found that the change of evolution equations is the main cause of the reduction in uncertainty\footnote{While the flavor nonsinglet evolution equation \eqref{NS_evol} remains unchanged, our choice to put $N_f =0$ still has an effect on the initial conditions for the flavor nonsinglet dipole amplitudes since the SIDIS data are sensitive to both the flavor singlet and nonsinglet dipole amplitudes (see \eq{g1h}).} for $\Delta s^-$, and that the inclusion of $pp$ data constrains $\Delta s^+$ to be consistent with zero at $x \lesssim 0.1$ and $\Delta s^-$ to be negative as $x\to 0.1$.

\subsection{Matching onto DGLAP-based {\boldmath $\Delta g$} at larger $x$ and EIC impact study} \label{ssec:matching}

\subsubsection{Matching to DGLAP-based analysis}

\begin{figure}[b!]
\begin{centering}
\includegraphics[width=500 pt]{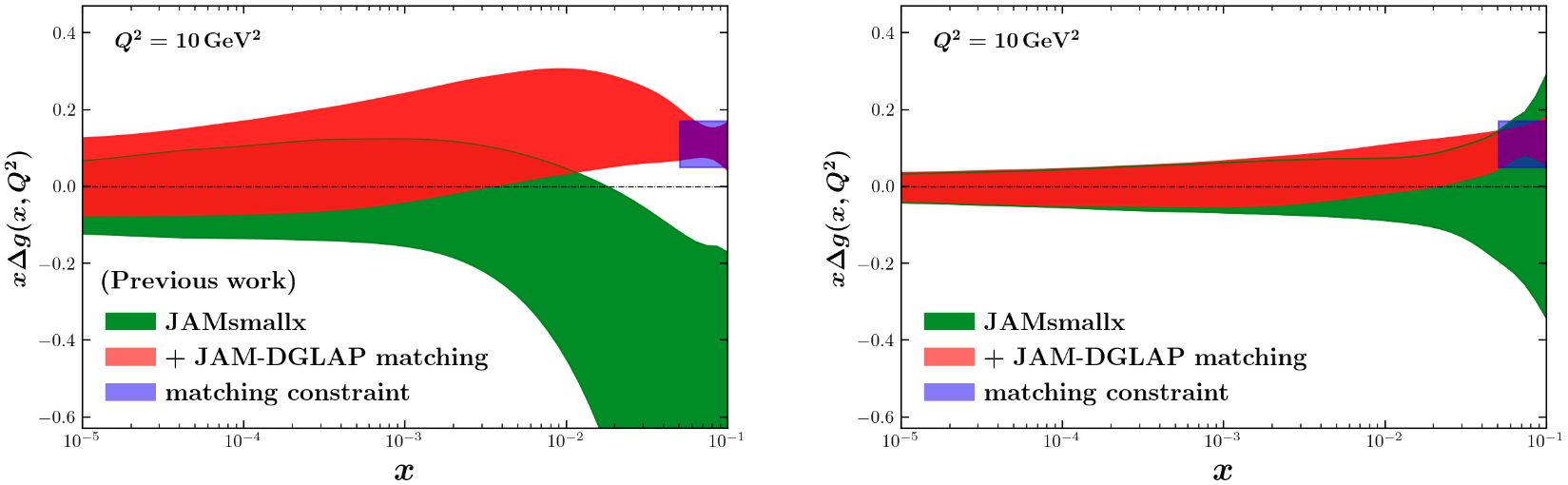}
\caption{$x\,\Delta g(x, Q^2)$ versus~$x$ with and without matching onto the JAM-DGLAP fits of Refs.~\cite{Zhou:2022wzm, Cocuzza:2022jye, Hunt-Smith:2024khs} (``SU(3)$+$positivity scenario", ``JAM", and ``$+$LQCD", respectively). The baseline JAMsmallx fit (green) is compared with the fit with matching (red), and the blue box is a union of all three JAM-DGLAP $\Delta g(x, Q^2)$ extractions \cite{Zhou:2022wzm,Cocuzza:2022jye, Hunt-Smith:2024khs} in the range $10^{-1.3} < x < 0.1$. The result of such a matching from our previous analysis \cite{Adamiak:2023yhz} (left) is compared with the current analysis (right).
    \label{DGLAP_DeltaG}
}
\end{centering}
\end{figure}

Eventually, the small-$x$ analysis here must interface with the DGLAP-based approach at larger $x$. 
Rigorously formulating and implementing such matching is rather involved, and is left for future work. 
Instead, to study the effect that the small-$x$/large-$x$ interface could have on the small-$x$ analysis, we performed a simple matching between the JAMsmallx and JAM-DGLAP gluon hPDFs (a union of ``SU(3)$+$positivity scenario", ``JAM",  and ``$+$LQCD" extractions from Refs.~\cite{Zhou:2022wzm}, \cite{Cocuzza:2022jye}, and \cite{Hunt-Smith:2024khs}, respectively) at $x \approx 0.1$. 
The results of the matching exercise are shown in \fig{DGLAP_DeltaG}, where we plot $x \, \Delta g (x, Q^2)$ from the previous~\cite{Adamiak:2023yhz} (left panel) and current (right panel) small-$x$ analyses with (in red) and without (in green) matching to the large-$x$ sector. 
The blue box encompasses the spread of all three gluon hPDF extractions from Refs.~\cite{Zhou:2022wzm, Cocuzza:2022jye, Hunt-Smith:2024khs} in a chosen large-$x$ range. 
The matching is accomplished by imposing a $\chi^2$ penalty on replicas that lie outside of the blue box, {\it i.e.}, outside the approximate range provided by the JAM-DGLAP extractions of $\Delta g(x, Q^2)$~\cite{Zhou:2022wzm, Cocuzza:2022jye, Hunt-Smith:2024khs}. 
In practice, we apply a penalty to replicas that fall outside $0.05 < \Delta g(x,Q^2=10~\mathrm{GeV}^2) < 0.17$ near the initial condition $x \approx 0.1$ (specifically, in the $x$-range of $10^{-1.3} < x < 0.1$). 
As it stands, the uncertainty band of our current JAMsmallx baseline fit already encompasses this range, as seen by the green band in the right panel of Fig.~\ref{DGLAP_DeltaG} overlapping the blue box.
The $\chi^2$ penalty then effectively restricts the spread of replicas (as seen by the red band) rather than shifting them outright like in Fig.~18 of Ref.~\cite{Adamiak:2023yhz} (reproduced in the left panel of Fig.~\ref{DGLAP_DeltaG}).

This comparison provides an optimistic outlook for the future work of continuously matching large-$x$ and small-$x$ extractions of hPDFs. 
The previous analysis with only DIS and SIDIS data~\cite{Adamiak:2023yhz} presented a fundamental incompatibility between the JAM-DGLAP and JAMsmallx results at the $x\approx 0.1$ interface (the green band in the left panel of \fig{DGLAP_DeltaG} does not overlap with the blue box), whereas the current analysis including the $pp$ data does not have this issue.
Note that the $0.01 \lesssim x < 0.1$ region of $\Delta g$ from JAM-DGLAP is clearly more positive than the JAMsmallx baseline fit, although this difference does not greatly affect the small-$x$ asymptotics.
The span of the uncertainty encompasses zero nearly evenly in the extrapolation region ($x< 0.005)$ even after matching, allowing for some replicas to still predict asymptotically negative $\Delta g$ values.

\begin{figure}[t] 
\begin{centering}
\includegraphics[width=500 pt]{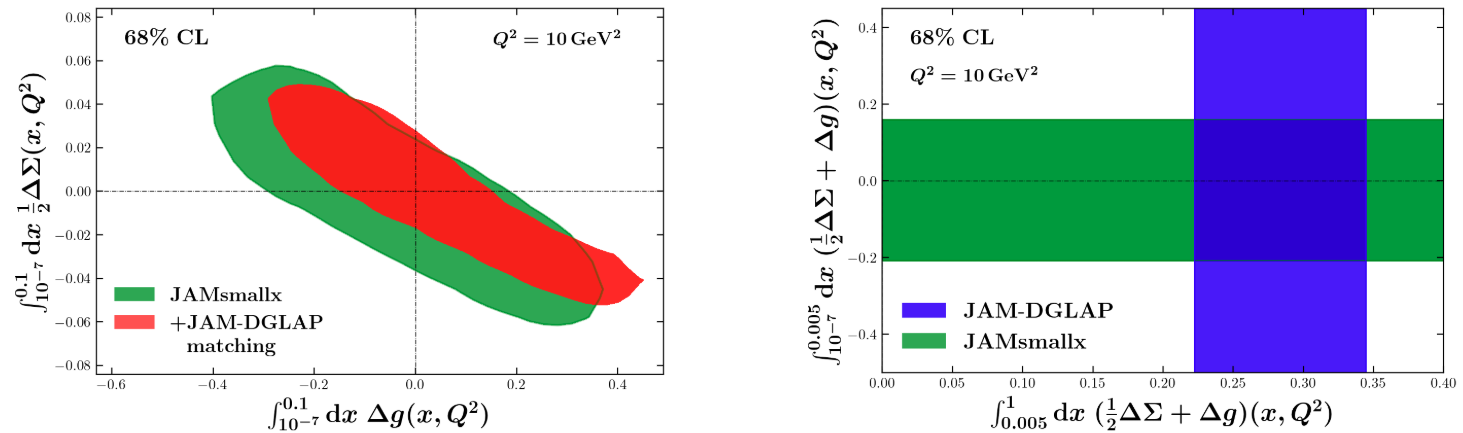}
\caption{(Left) A comparison of the $10^{-7}< x < 0.1$ truncated integrals of JAMsmallx analyses with (red) and without (green) matching onto JAM-DGLAP extractions.
(Right) An exclusion plot of two truncated integrals:~the vertical axis shows the truncated integral of total parton helicity over the extrapolation/prediction range of $10^{-7}< x < 0.005$, while the horizontal axis shows the truncated integral of total parton helicity over the data-driven range $0.005< x< 1$. The JAMsmallx analysis (green) does not fit data for $x > 0.1$, 
while the JAM-DGLAP analysis ~\cite{Cocuzza:2022jye} (blue) has uncontrolled extrapolation for $x < 0.005$.
The overlapping region is the best prediction for total parton helicity that simultaneously satisfies both analyses. 
    \label{DGLAP_trunc}
}
\end{centering}
\end{figure}

In the left panel of Fig.~\ref{DGLAP_trunc} we show how the small-$x$ parton helicity contributions have been affected by this matching, where 
we plot $\Delta g$ and $\tfrac{1}{2}\Delta\Sigma$ integrated over the range $x\in[10^{-7}, 0.1]$ at $Q^2 = 10$~GeV$^2$ on the horizontal and vertical axes, respectively. 
The integrated small-$x$ gluon and total helicity contributions of the fit with the JAM-DGLAP matching  at $Q^2 = 10~\mathrm{GeV}^2$ now trend positive due to the large amount of negative $\Delta g(x, Q^2)$ replicas removed in the $0.01 \lesssim x < 0.1$ range,
\begin{equation}
    \int\limits_{10^{-7}}^{0.1}\,\dd x \, \Delta g (x, Q^2)\Big|_{+\,{\rm matching}} \approx 0.08 \pm 0.24 \,, \qquad \int\limits_{10^{-7}}^{0.1}\,\dd x \,  \Bigl(\frac{1}{2}\Delta\Sigma + \Delta g \Bigr) (x, Q^2)\Big|_{+\,{\rm matching}} \approx 0.08 \pm 0.21\;.
\end{equation}
This finding should be taken with the caveat that a full DGLAP-matching procedure would require a composite small-to-large-$x$ formalism to simultaneously satisfy DGLAP large-$x$ and KPS-CTT small-$x$ analyses.

In the right panel of Fig.~\ref{DGLAP_trunc} we show 
truncated integrals of the total parton helicity for JAMsmallx over the extrapolation/prediction region $x\in[10^{-7}, 0.005]$ (plotted on the vertical axis) as well as JAM-DGLAP over the data-driven range $x\in[0.005, 1]$ (plotted on the horizontal axis).  We have verified that the truncated integrals of both analyses are consistent with each other over the shared region $0.005 \leq x \leq 0.1$ where both are constrained by data.  
This calculation allows us to estimate for the first time the total parton helicity contribution over an extended small-to-large range of $x$. Recall that JAMsmallx analyses do not include data in the range $x>0.1$; likewise, JAM-DGLAP analyses are not constrained in the extrapolation region $x < 0.005$ (where there are no experimental data). The 68\% confidence level bands for these two analyses are only trustworthy in their respective regions of $x$, and are simply plotted across the entire vertical (for JAM-DGLAP) or horizontal (for JAMsmallx) axes. The region of overlap indicates which truncated integrals of the total parton helicity satisfy both the JAMsmallx and JAM-DGLAP analyses. The lower-left and upper-right corners of the overlapping region then act as the lower and upper boundaries of the 68\% confidence level for a ``full-$x$" (or, rather, over the range $10^{-7} < x < 1$) total parton helicity contribution to the proton spin, for which we find at $Q^2 = 10~\mathrm{GeV}^2$
\begin{equation}\label{extended_truncation}
    \int\limits_{10^{-7}}^{1}\,\dd x \,  \Bigl(\frac{1}{2}\Delta\Sigma + \Delta g \Bigr) (x, Q^2)\ \in\ [0.02, 0.51].
\end{equation}
Of course the extended range $10^{-7} < x < 1$ still does not cover the entire region of $x\in [0,1]$.  However, parton saturation will dominate at extremely small $x < 10^{-7}$ (at $Q^2 = 10$~GeV$^2$), and our expectation, based on the similar effect in the Reggeon evolution \cite{Itakura:2003jp}, is that these effects will make the parton helicity and orbital angular momentum (OAM) negligible in that region.  If that is the case, the truncated integral in Eq.~\eqref{extended_truncation} is a reasonable estimate, at present, for the total parton helicity contribution to the proton spin sum rule~\eqref{spin_sum}.  This would imply that parton OAM contributes between $-0.01$ and $0.48$ to the proton spin, within a 68\% confidence interval.

\subsubsection{EIC impact study}

In this subsection we report the results of an EIC impact study, similar to that performed in Ref.~\cite{Adamiak:2023yhz}, but now using our DIS+SIDIS+$pp$ small-$x$ analysis as the baseline.
The EIC pseudodata is generated for the polarized DIS and SIDIS processes in the kinematic ranges $10^{-4} < x < 0.1$ and $1.69~\mathrm{GeV}^2 < Q^2 < 50~\mathrm{GeV}^2$, with a 2\% point-to-point uncorrelated systematic uncertainty as given from the EIC Yellow Report~\cite{AbdulKhalek:2021gbh, VanHulse:2023uga}.
\footnote{The CM energies for DIS pseudodata on a proton were at $\sqrt{s} = \{29, 45, 63, 141\}~\mathrm{GeV}$ with the integrated luminosity of $100~\mathrm{fb}^{-1}$. For DIS on a deuteron and ${}^3\mathrm{He}$, the energies were $\sqrt{s} = \{29,66,89\}~\mathrm{GeV}$ with the integrated luminosity of $10~\mathrm{fb}^{-1}$. For SIDIS on a proton the CM energy was $\sqrt{s} = 141~\mathrm{GeV}$ with the integrated luminosity of $10~\mathrm{fb}^{-1}$.} 
As discussed previously, our small-$x$ prediction for the $g_1^p(x, Q^2)$ structure function differs from that in Ref.~\cite{Adamiak:2023yhz} when simultaneously fitting polarized $pp$ data. 
Therefore, the EIC pseudodata for DIS and SIDIS are also different here compared to Ref.~\cite{Adamiak:2023yhz}. 
We generate three different sets of EIC pseudodata using the hPDF central curves from three different scenarios:~all replicas, only the replicas with asymptotically positive $g_1^p(x, Q^2)$ as $x \to 0$, and only the replicas with asymptotically negative $g_1^p(x, Q^2)$ as $x \to 0$. 
We then ran three new global analyses that simultaneously include the existing experimental DIS, SIDIS, and $pp$ data, along with one of the aforementioned sets of EIC pseudodata.

\begin{figure}[t] 
\begin{centering}
\includegraphics[width=510 pt]{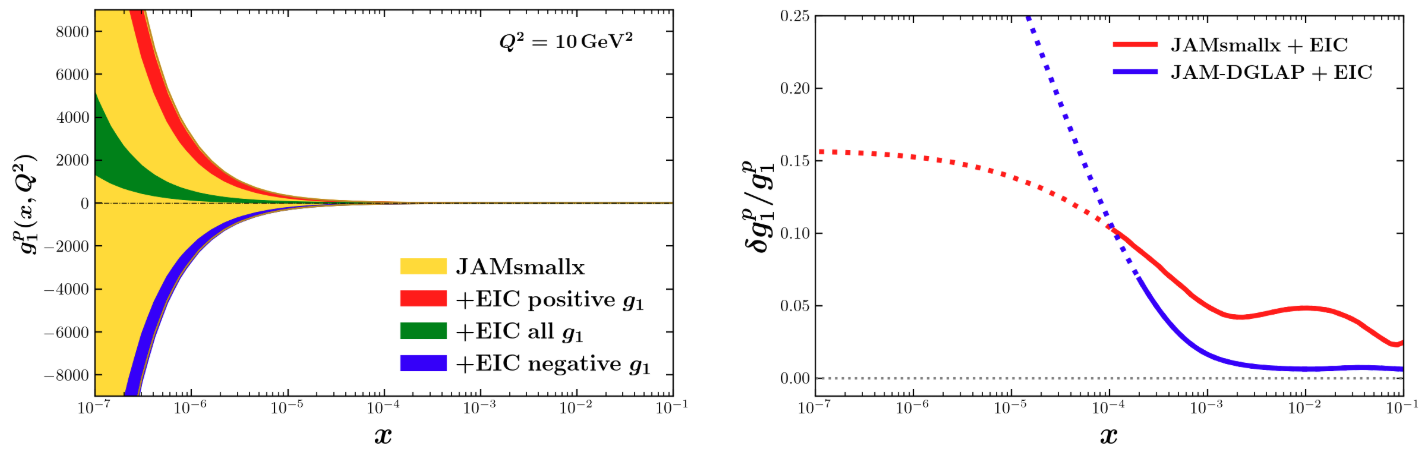}
\caption{(Left) EIC impact on $g_1^p(x, Q^2)$ at $Q^2=10~\mathrm{GeV}^2$ for the three scenarios described in the text (red, green, blue) compared to the JAMsmallx global analysis of current DIS, SIDIS, and $pp$ experimental data (yellow).
(Right) Relative uncertainties for the EIC impact study conducted here using small-$x$ helicity evolution (red) compared to one using DGLAP evolution (blue). All results are for the asymptotically positive $g_1^p$ scenario.  The lines become dotted in the region of $x$ below which the pseudodata was generated:~the JAMsmallx analysis generated pseudodata down to $x = 10^{-4}$, while the JAM-DGLAP analysis generated pseudodata down to $x = 2\times10^{-4}$ \cite{Zhou:2021llj}.
    \label{EIC_impact}
}
\end{centering}
\end{figure}

In Fig.~\ref{EIC_impact} we show the impact that these various EIC pseudodata sets will have on our small-$x$ uncertainties of the $g_1^p$ structure function of the proton. The left panel shows the uncertainty bands for $g_1^p(x,Q^2)$ (at $Q^2 = 10\,{\rm GeV^2}$) as a function of $x$, while the right panel compares the relative uncertainty (the ratio of $g_1^p$'s uncertainty to its magnitude) of a JAMsmallx fit to a JAM-DGLAP fit when both are additionally including EIC pseudodata. Keeping in mind the small-$x$ uncertainty has already been drastically reduced by the inclusion of $pp$ data alone, it is clear from the left panel of Fig.~\ref{EIC_impact} that data at even lower $x$ from the EIC will further decrease our uncertainties significantly. The uncertainty band at $x = 10^{-7}$ for the $g_1^p(x,Q^2)$ extraction when fitting to EIC pseudodata generated from the mean of all replicas (green band in the left panel of Fig.~\ref{EIC_impact}) is reduced by 93\% compared to the JAMsmallx baseline fit (yellow band). This reduction is in addition to the 70\% reduction quoted earlier due to our inclusion of the polarized $pp$ data, which highlights the importance of $pp$ data in maximizing the constraints on $g_1^p(x,Q^2)$, even with the very precise DIS and SIDIS measurements that the EIC will provide. The bimodal split of asymptotically positive or negative solutions can also seemingly be resolved via EIC data, provided its measurements of $g_1^p(x,Q^2)$ are slightly positive or negative in the range $10^{-4} \lesssim x \lesssim 5\times 10^{-3}$ (see the left panel of Fig.~\ref{g1_band_comparison}). The right panel of Fig.~\ref{EIC_impact} reinforces the advantage of small-$x$ helicity evolution, namely that it has predictive power for the hPDFs and observables at small $x$ (beyond where there are experimental data), whereas for DGLAP-based analyses it is an {\sl ad hoc} extrapolation into that region. Comparing the relative uncertainties of the $g_1^p$ structure function from JAMsmallx and a JAM-DGLAP~\cite{Zhou:2021llj} extraction, we find that in the region beyond where EIC pseudodata was generated, the small-$x$ helicity evolution equations give significantly smaller and more stable uncertainties.

\section{Conclusions} \label{sec:conclusions}

This paper details the first simultaneous analysis of polarized DIS, SIDIS, and proton-proton collision data in the framework of the KPS-CTT small-$x$ helicity evolution equations~\cite{Kovchegov:2015pbl, Kovchegov:2018znm, Cougoulic:2022gbk}. Using the small-$x$ $gg \to g$ particle production channel in polarized $pp$ collisions~\cite{kovchegov:2024aus}, we incorporated $A_{LL}^{\rm jet}$ data into our previous DIS+SIDIS only analysis~\cite{Adamiak:2023yhz}. To account for only gluons being present in the $pp$ process, we also modified the large-$N_c\&N_f$ evolution equations by setting $N_f=0$ into what we dubbed the ``large-$N_c^{+q}$" evolution equations. Using the analytical results in Eqs.~(\ref{pJet_GA}), (\ref{hPDFs_log}), (\ref{g1_log}), the initial conditions~(\ref{Dipole_ICs}), and the evolution equations~(\ref{eq_LargeNc+q}), our global analysis shows compatibility between polarized DIS and SIDIS measurements and polarized $pp$ jet production data at small $x< 0.1$ ($N_{\rm pts}=240$), with $\chi^2/N_{\mathrm{pts}} = 0.98$. We found that the inclusion of a gluon-sensitive polarized $pp$ observable does indeed better constrain the gluon-dominated polarized dipole amplitudes $\widetilde{G}$ and $G_2$, leading to an approximately 70\% reduction in the small-$x$ theoretical uncertainty for $\Delta g(x,Q^2)$, $\Delta\Sigma(x,Q^2)$, and $g_1^p(x,Q^2)$ at $Q^2 = 10~\mathrm{GeV}^2$ (Figs.~\ref{DeltaG_DeltaSigma} and \ref{g1_band_comparison}). In addition, the $g_1^p$ replicas are now equally distributed between asymptotically positive and negative solutions at small $x$, and there is a significant reduction in replicas with $\Delta g < 0$ for $0.01 \lesssim x< 0.1$.

Moreover, the constraining power of polarized $pp$ data has changed the predictions for helicity at small $x$ quite significantly (Fig.~\ref{xmin_integral_comparison}), due to the fact that the gluon hPDF spans zero evenly across the entire range $x< 0.1$ and its small-$x$ uncertainty has been substantially reduced (Fig.~\ref{DeltaG_DeltaSigma}, left panel). A salient feature from Ref.~\cite{Adamiak:2023yhz} was the prediction that the total helicity contribution from quarks and gluons at small $x$ was {\it negative} within $1\sigma$ uncertainty, leading to the conclusion that there may be a significant contribution needed from parton orbital angular momentum to satisfy the Jaffe-Manohar spin sum rule. After including polarized $pp$ data in the current analysis, we obtained a much better constraint on the gluon hPDF $\Delta g$, resulting in the function having moved much closer to zero at $0.01\lesssim x< 0.1$. While only $\Delta g$ is \textit{directly} sensitive to $pp$ data, the constraints from those measurements on the initial conditions have an effect on all polarized dipole amplitudes, and thus hPDFs, through evolution. We find that the flavor singlet combination of the quark hPDFs, $\Delta\Sigma$, has also shifted to be more consistent with zero at small~$x$ (Fig.~\ref{DeltaG_DeltaSigma}, right panel).
Summarizing these observations, we found that polarized $pp$ data cause the total parton helicity at small $x$ to be consistent with zero, although there are still large uncertainties of $\pm \sim 0.25$. Through an EIC impact study, we also found, as in Ref.~\cite{Adamiak:2023yhz}, that the asymptotic $x\to 0$ behavior of $g_1^p$ (at $Q^2 = 10~\mathrm{GeV}^2$) can be determined rather accurately once polarized DIS/SIDIS data for $10^{-4} \lesssim x < 5\times 10^{-3}$ becomes available.

Combining our JAMsmallx and a recent JAM-DGLAP extraction, we also estimated the total quark and gluon helicity contribution from $10^{-7} < x < 1$.
For $x < 0.005$, JAM-DGLAP is not reliably constrained by data, whereas the predictive power of JAMsmallx given by the small-$x$ helicity evolution equations makes the statistical uncertainties more trustworthy in this region. 
On the other hand, JAMsmallx extractions are restricted to $x < 0.1$ data, whereas JAM-DGLAP includes data over the range $0.005 < x < 1$.
Using the overlap region of the 68\% confidence level bands for the small-$x$ ($10^{-7} < x < 0.005$) truncated integral of $\Delta g(x,Q^2) + \frac{1}{2}\Delta\Sigma(x,Q^2)$ from JAMsmallx and the data-driven ($0.005 < x < 1$) truncated integral from JAM-DGLAP yields a total parton helicity for $x > 10^{-7}$ of between 0.02 and 0.51.
If in addition we assume that parton saturation will make parton helicity and OAM in the region of $x < 10^{-7}$ numerically negligible, we may conclude that the range $[0.02,0.51]$ is a reasonable estimate, at present, for the total parton helicity contribution to the proton spin sum rule~\eqref{spin_sum}.  This implies, within a 68\% confidence interval, that parton OAM contributes between $-0.01$ and $0.48$ to the proton spin.

In the future, the theoretical calculation of the remaining parton production channels in small-$x$ polarized $pp$ collisions, while a difficult task, is still needed for a more complete phenomenological analysis. The remaining channels may indeed be sensitive to the other polarized quark dipole amplitudes, giving additional constraints on the hPDFs. A complete parton production cross section in polarized $pp$ collisions would also allow us to re-institute using the more accurate large-$N_c\&N_f$ evolution equations, with the most updated DLA evolution equations from Ref.~\cite{Borden:2024bxa}. This would further set the stage to include other observables measured in longitudinally polarized proton-proton collisions, such as single-inclusive hadron, isolated direct photon, and weak gauge boson production into our global analysis (see Ref.~\cite{RHICSPIN:2023zxx} for a summary of the Relativistic Heavy Ion Collider $A_{LL}$ data).

On a somewhat longer time scale, one may also attempt to describe the data on $A_{LL}$ of dijets produced in polarized $pp$ collisions \cite{STAR:2021mqa} in the small-$x$ framework. Furthermore, there is an avenue to improve the accuracy of the polarized $pp$ jet production cross section in Eq.~\eqref{num_xperp} by incorporating higher-order $\ln(R)$ corrections to the jet function. Inclusive jet spectra have been shown to have large radius-dependent deviations even at LO, which would introduce dependence on $z_j$, the transverse momentum fraction of the parton carried by the produced jet~\cite{Kaufmann:2014nda, Dasgupta:2014yra, Kang:2016mcy}. Bringing in $\ln(R)$ corrections can be done separately from incorporating the remaining parton production channels and may improve the fits to jet production data. We conclude that there are several promising pathways to continue developing the modern theory and phenomenology of spin at small $x$.

\section*{Acknowledgments}\label{sec:acknowledgement}

We thank Trey Anderson and Chris Cocuzza for providing replicas and calculations for the JAM-DGLAP-based analyses. This work has been supported by the U.S. Department of Energy, Office of Science, Office of Nuclear Physics under Award Number DE-SC0004286 (YK and ML), No. DE-SC0020081 (AT), No. DE-SC0024560 (MS), No.~DE-AC05-06OR23177 (DA, WM and NS) under which Jefferson Science Associates, LLC, manages and operates Jefferson Lab, and the National Science Foundation under Grant No.~PHY-2308567 (DP). The work of NS was supported by the DOE, Office of Science, Office of Nuclear Physics in the Early Career Program, and the work of MS was supported by the DOE, Office of Science, Office of Nuclear Physics under the Funding for Accelerated, Inclusive Research Program. This work is also supported by the U.S. Department of Energy, Office of Science, Office of Nuclear Physics, within the framework of the Saturated Glue (SURGE) Topical Theory Collaboration. The work of YT has been supported by the Research Council of Finland, the Centre of Excellence in Quark Matter and projects 338263, 346567 and 359902, and by the European Research Council (ERC, grant agreements No. ERC-2023-COG-101123801 GlueSatLight and No. ERC-2018-ADG-835105 YoctoLHC). The content of this article does not reflect the official opinion of the European Union and responsibility for the information and views expressed therein lies entirely with the authors.


%

\end{document}